\documentclass[aps,pra,reprint,superscriptaddress,longbibliography]{revtex4-2}
\usepackage{graphicx}
\usepackage{amssymb}
\usepackage{subfigure}
\usepackage{amsmath}
\usepackage{amsfonts}
\usepackage{bm}
\usepackage{color}
\usepackage{array}
\usepackage{rotating}
\usepackage{multirow}
\usepackage[hidelinks]{hyperref}
\usepackage{url}
\usepackage{lineno}
\usepackage[utf8]{inputenc}

\begin{document}


\title{Discretized continuous quantum-mechanical observables are neither continuous nor discrete}


\author{Thais L. Silva}
\email{thaisdelimasilva@gmail.com}
\affiliation{Instituto de F\'isica, Universidade Federal do Rio de Janeiro, Caixa
Postal 68528, Rio de Janeiro, RJ 21941-972, Brazil}

\author{\L ukasz Rudnicki}

\affiliation{International Centre for Theory of Quantum Technologies (ICTQT), University of Gda{\'n}sk, 80-308 Gda{\'n}sk, Poland}
\affiliation{Center for Theoretical Physics, Polish Academy of Sciences, Al.
Lotnik\'ow 32/46, 02-668 Warsaw, Poland}

\author{Daniel S. Tasca}
\affiliation{Instituto de F\'isica, Universidade Federal Fluminense, Niter\'oi, RJ
24210-346, Brazil}

\author{Stephen P. Walborn}
\affiliation{Instituto de F\'isica, Universidade Federal do Rio de Janeiro, Caixa
Postal 68528, Rio de Janeiro, RJ 21941-972, Brazil}
\affiliation{Departamento de F\'{\i}sica, Universidad de Concepci\'on, 160-C Concepci\'on, Chile}
\affiliation{ANID – Millennium Science Initiative Program – Millennium Institute for Research in Optics, Universidad de Concepci\'on, 160-C Concepci\'on, Chile}
\begin{abstract}
Most of the fundamental characteristics of quantum mechanics, such as non-locality and contextuality, are manifest in discrete, finite-dimensional systems.   However, many quantum information tasks that exploit these properties cannot be directly adapted to continuous variable systems.  To access these quantum features, continuous quantum variables can be made discrete by binning together their different values, resulting in observables with a finite number ``$d$" of outcomes.  While direct measurement indeed confirms their manifestly discrete character, here we employ a salient feature of quantum physics known as mutual unbiasedness to show that such coarse-grained observables are in a sense neither continuous nor discrete. Depending on $d$, the observables can reproduce either the discrete or the continuous behavior, or neither. To illustrate these results, we present an example for the construction of such measurements and employ it in an optical experiment confirming the existence of four mutually unbiased measurements with $d = 3$ outcomes in a continuous variable system, surpassing the number of mutually unbiased continuous variable observables.  
\end{abstract}



\maketitle

\par
\section{Introduction}
  A multitude of quantum information protocols have been developed that exploit fundamental aspects of discrete quantum systems, such as non-locality and contextuality.  However, many systems are described by continuous or infinite dimensional variables.  A continuous system is fundamentally different than a discrete one, and many quantum information tasks that have been designed for discrete systems cannot be directly adapted to continuous variable systems.  To overcome this, many authors have “discretized” the continuous variable states and/or observables, an approach that has been quite successful in terms of translating protocols that were originally cast in the discrete regime into the continuous realm.   For example, it is well known that it is difficult to employ the usual phase space operators, such as quadratures of position and momentum, to demonstrate quantum non-locality \cite{revzen05}.  This has led to a number of binning schemes applied to either the measurements or the states \cite{gilchrist98,banaszek98,banaszek99,wenger03,cavalcanti11,ketterer15}.  Similar types of discretization have been applied to quantum contextuality \cite{massar01,plastino10,asadian15,finot17}, the quantum search algorithm \cite{ketterer14}, entanglement detection \cite{Gneiting11,Carvalho12,Tasca18b} and general approaches to realize finite-dimensional quantum information protocols in continuous variable systems \cite{Vernaz-Gris14,Ketterer16}.  

It may be tempting to consider these “discretized continuous observables” as completely analogous to discrete ones, as they can be made to reproduce many of the key features.  For example, clearly any direct measurement of such observables should trivially result in discrete probability distributions describing the $d$ measurement outcomes.  However, when measuring different observables in distinct phase space directions (all encoded in discrete probabilities) it might occur that correlations between the results exhibit a behavior that is characteristic of a CV system. \textit{A priori}, it is not clear whether a discrete and finite-dimensional (finite number of bins) coarse graining of continuous variables (CV) does fully behave in a ``discrete-like" manner, or rather preserves some of the features of the underlying CV system. 
\par
One of these quantum features is mutual unbiasedness, where eigenstates of one measurement give equally likely outcomes for a second measurement, and which displays a key distinction between discrete and continuous systems.  Namely, in the continuous case, there can be at most three mutually unbiased observables\cite{Weigert08}, while for discrete systems there can be up to $d+1$ for particular values of the dimension $d$ (when $d$ is the power of a prime number)  \cite{Ivonovic1981,wootters1989,Bandyopadhyay2002,klappenecker2003}. 
\par
The distinction between the continuous and discrete cases suggests that mutual unbiasedness can be used as a benchmark for the continuous/discrete nature of coarse-grained observables of CV systems.
In the present manuscript we pursue this approach and discover a surprising result:  in general, the observables are neither continuous nor discrete.  That is, depending on the number of possible measurement outcomes, they can reproduce the discrete behavior or the continuous behavior, or give definite results that correspond to neither.  
\par
Let us briefly summarize the results we present below. 
We consider periodic discretization with $d$ bins and we show that the maximum number of mutually unbiased observables is $R_{\mathrm{max}} = d + 1$ only for prime $d$. This is also an exclusive instance which fully reproduces the discrete pattern, as in all other cases (including power-prime dimensions) the situation is more complex. However, the rule which applies in general tells us that for even dimension d, we get $R_{\mathrm{max}} = 3$, imitating the CV case. Thus, surprisingly, we find that the investigated observables are neither continuous nor discrete, in what concerns mutual unbiasedness, at the same time showing a certain level of compatibility with both domains.
An interesting aspect of our coarse graining approach is that it allows for the derivation of explicit results for any  dimension parameter $d$, not just for  prime power numbers, a feature yet to be achieved for discrete variable systems.  We provide an accessible proof of these theoretical results, as well as an experimental demonstration of four mutually unbiased observables in a continuous system for the first time.   On the one hand, our results show that discretization of continuous variables, though it may reproduce many key features of discrete systems, can also lead to unique properties that are neither continuous nor discrete.  On the other hand, given an $R_{\mathrm{max}}$ we can find a $d$ which allows the construction of $R_{\mathrm{max}}$ mutually unbiased observables. Thus, our work shows that it is possible to identify an arbitrary number of mutually unbiased observables in a continuous quantum system.  This can have important applications in quantum tomography and randomness generation, as well as other applications.       

\begin{figure}
\includegraphics[width=8cm]{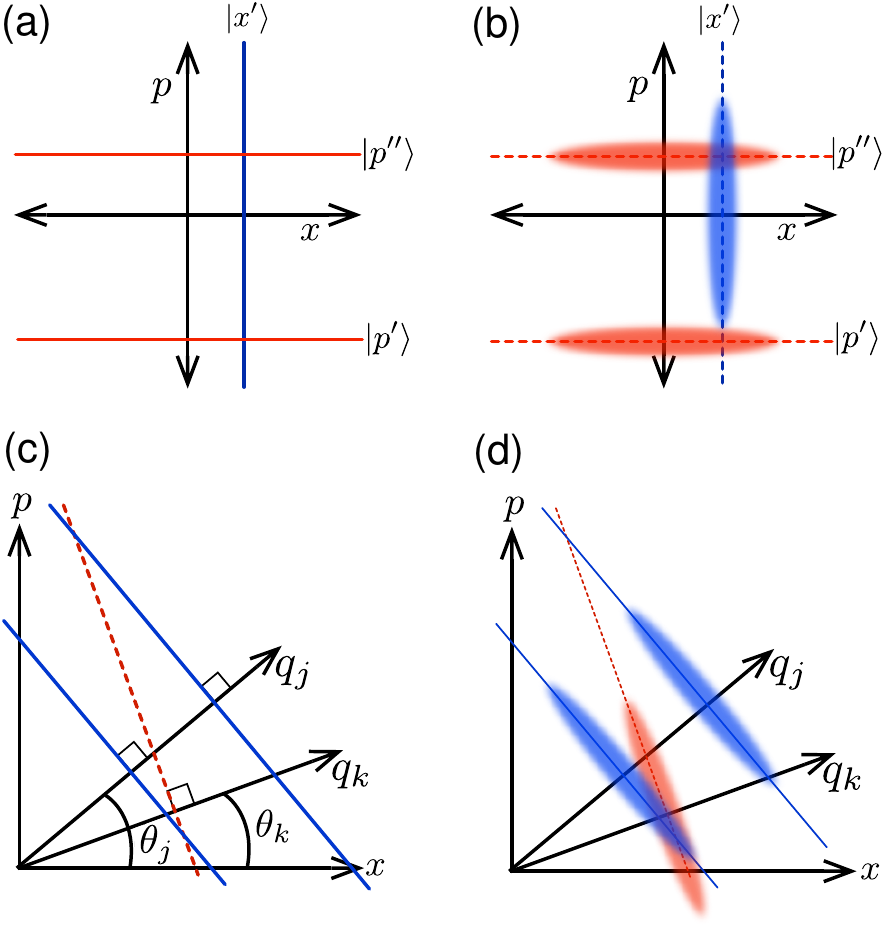}
\caption{Pictorial illustration of mutual unbiasedness in phase space.  a) Complementary position and momentum eigenstates are  distributions denoted by the vertical and horizontal lines, respectively.  The overlap between any position eigenstate and any momentum eigenstate is the same, represented by the intersection of the distributions. b) The non-physical eigenstates maintaining complementarity are approximated by physical states that are given by positive distributions centered around the eigenstates, with finite widths in both phase-space directions.  The overlap between physical states is represented graphically by the overlap of the distributions.  The physical states are no longer mutually unbiased. c) Eigenstates of rotated observables $\hat{q}_j$ and $\hat{q}_k$, illustrating that any two non-parallel observables are mutually unbiased, as per Eq. \eqref{eq:CVs}. d) Physical states corresponding to the eigenstates in c) also lose their mutual unbiasedness. } 
\label{fig:vecs1}
\end{figure}
\section{Mutual unbiasedness}
 The original treatment \cite{Schwinger60,Kraus87} of mutual unbiasedness concerns mutually unbiased bases (MUB), namely, sets of rank $1$ projectors such that the overlaps between all projectors from different sets (overlaps of elements from different bases) are equal --- implying that full information according to a given basis means no information with respect to all other bases \cite{Durt10}. As already mentioned, discrete and finite $d$-dimensional quantum systems admit at most $d+1$ MUBs \cite{Ivonovic1981,wootters1989,Bandyopadhyay2002,klappenecker2003}, 
%
while there are at most three MUBs for a CV system  \cite{Weigert08}.  Indeed, most quantum mechanics textbooks highlight the fact that mutual unbiasedness between position ($\hat{x}$) and momentum ($\hat{p}$) operators can be demonstrated by the eigenstate projections $|\langle x | p\rangle| = 1/\sqrt{2 \pi}$ (we set $\hbar$ = 1 throughout).  What is somewhat less well-known is the fact that \emph{any} two non-parallel phase space operators $\hat{q}_j=\cos \theta_j \hat{x} + \sin \theta_j \hat{p}$ and $\hat{q}_k$ (defined analogously) are mutually unbiased (see Fig. \ref{fig:vecs1} c), i.e.
\begin{equation}
|\langle q_j| q_k\rangle| = \left({2 \pi |\sin \theta_{jk}}|\right)^{-1/2},
\label{eq:CVs}
\end{equation}
 where  $\theta_{jk} \equiv \theta_j - \theta_k\neq \{0,\pi\}$ is the angle between them \footnote{We note that in the limit $\theta_{jk} \rightarrow 0$, the limit must be taken before the absolute value to recover the normalization to the usual Dirac delta function.}, as illustrated in Fig. \ref{fig:vecs1}.  The MUBs stemming from three phase space operators can be achieved by selecting the same relative angle $\theta_{jk}=2\pi/3$, such that all pairs of operators satisfy Eq. \eqref{eq:CVs} with the same right-hand side \cite{Weigert08}.   Physical constraints on the eigenstates of  $\hat{q}_j$ and $\hat{q}_k$ also apply, destroying the mutual unbiasedness of the corresponding physical states. In real-world experiments, they are approximated by states that are localized around some mean value (see Fig. \ref{fig:vecs1}).    In addition, measurements in any quantum system suffer from some amount of coarse graining, which follows from the fact that every device has  finite resolution.  This can lead to practical consequences, such as overestimation of entanglement and/or security of quantum key distribution \cite{tasca13,Ray13a,Ray13b}.
Thus, mutual unbiasedness in the continuous variable regime occurs only for unphysical states, or requires consideration of measurements other than rank 1 projectors.  
\par
  There is no unique way for extending the notion of MUBs to projective measurements of rank higher than 1 and to POVMs in general. In finite dimension, one alternative is offered by a a family of POVMs \cite{Kalev14} labeled by the so-called efficiency parameter. However, these POVMs are only projective for maximal efficiency, when they boil down to rank 1 MUBs. On the other hand, every coarse graining of CV systems can be described by projective measurements ({with rank $\geq 1$)}. 

Being aware of such a discrepancy, we direct ourselves towards an experimentally-oriented view on mutual unbiasedness. In particular, we resort to an operational definition \cite{Tasca18a,paul18}, which says that a number $R$ of
sets (labeled by $j$) of $d$ projective measurement operators, $\left\{\hat{\Omega}_{j}^{(u)}\right\}_{u=0,...,d-1}$,  are mutually unbiased if for all $u_{0},\,u,\,v=0,\ldots,d-1$
\begin{equation}
p_{j}^{(u)}=\delta_{u,u_{0}}\;\Longrightarrow\;p_{j'}^{(v)}=d^{-1},\qquad j\neq j'.\label{MUB1}
\end{equation}
By $p_{j}^{(u)}$ we denote the probability associated with a direct measurement of $\hat{\Omega}_{j}^{(u)}$.
The aim of this definition is to convey the most important property of the original MUBs, namely, that localization with respect to one set implies even spreading from the perspective of all other sets.  We note in passing that this definition has recently been utilized for device independent applications \cite{kaniewski19}.

\subsection{Mutual unbiasedness through periodic coarse-graining}
Observables of continuous quantum variables can be discretized by a ``binning" procedure. One possibility is to divide the Hilbert space into pieces of finite size (e.g. intervals). To cover the whole range of values one needs infinitely many intervals of such type.  Another option stems from dividing the space into a \textit{finite} number of discrete parts. 
It has been shown that one path to theoretical and practical mutual unbiasedness in continuous variable systems as defined in Eq. \eqref{MUB1} is through  periodic coarse grained (PCG) observables. With this approach, physical mutually unbiased measurement pairs  \cite{Tasca18a} and mutually unbiased measurement triples \cite{paul18} have been demonstrated theoretically and experimentally.      
In this formulation, there naturally appears a ``dimensionality" parameter $d$, given by the number of possible measurement outcomes.   

\par
\begin{figure}
\includegraphics[width=8cm]{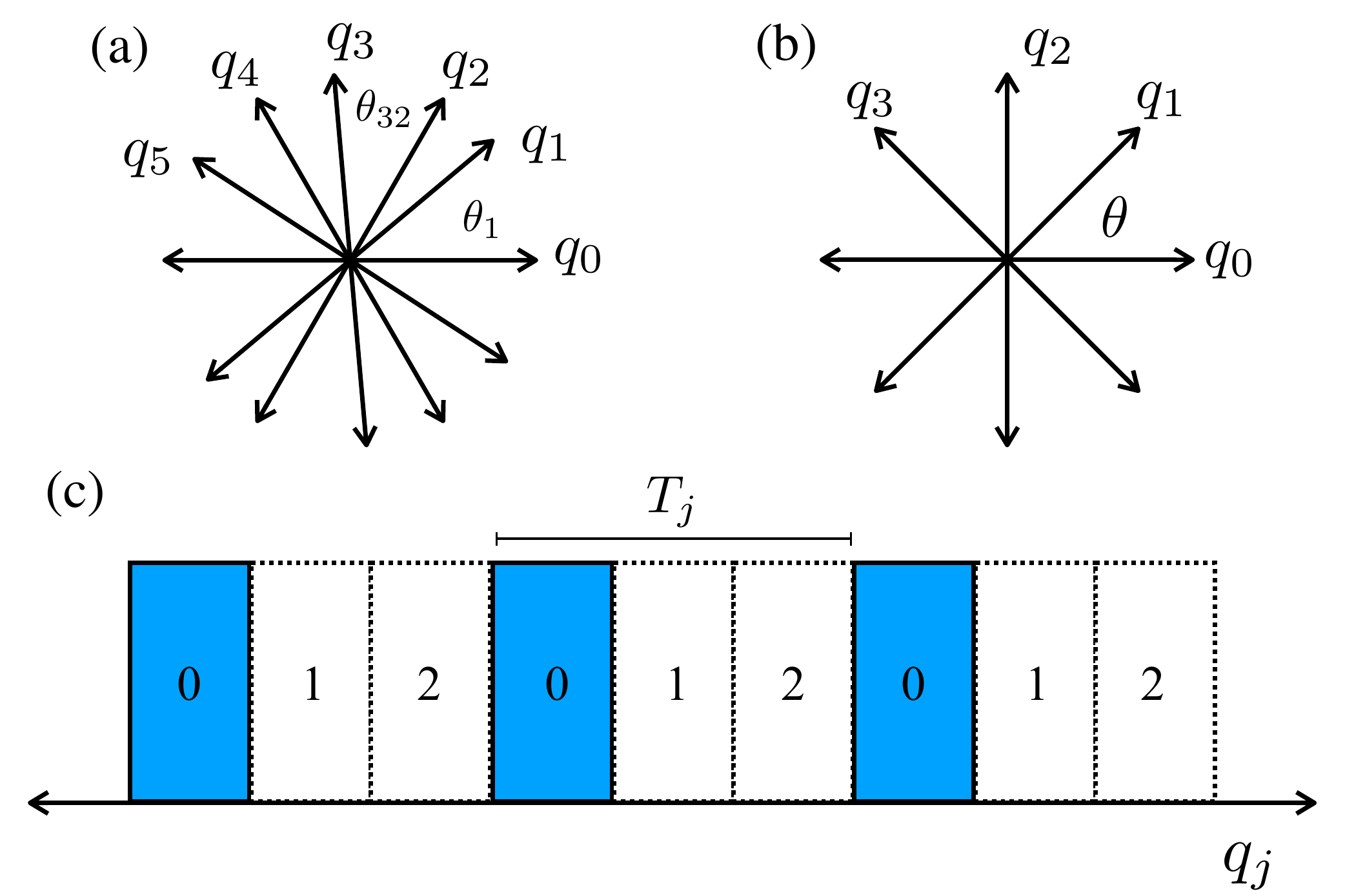}
\caption{a)  Examples of phase space variables for a) $R=6$ and  b) the $R=4$ implemented experimentally. Here $q_0$ is taken to be the position variable $x$. c) the coarse graining bin functions (experimentally implemented as amplitude masks) $M_{j,0}$ for $d=3$, represented by shaded region.}
\label{fig:vecs}
\end{figure}

Following Refs. \cite{Tasca18a,paul18}, let us consider $R$ phase space operators $\hat{q}_j$ as in Eq. \eqref{eq:CVs}, related to each other via phase space rotations, each characterized by an angle $\theta_j$, for $j=0,...,R-1$, as illustrated in Fig. \ref{fig:vecs} a).  We can then define $d$ coarse-grained projective measurement operators ($u=0,\ldots,d-1$):
 
\begin{equation}
\hat{\Omega}_{j}^{(u)}=\int dq_{j}\,M_{j,u}\left(q_{j}-q_{j}^{\textrm{cen}};T_{j}\right)\left|q_{j}\right\rangle \left\langle q_{j}\right|,
\label{Projectors}
\end{equation}
where the infinite set of rank one projectors $\left|q_{j}\right\rangle \left\langle q_{j}\right|$ are grouped according to the $d$ non-negative  ``bin functions'' $M_{j,u}$,
such that $\sum_{u=0}^{d-1}M_{j,u}=1$. The displacement
parameter $q_{j}^{\textrm{cen}}$ is included to allow for
freedom to define the origin, and the parameter $T_{j}$ is
the period of the bin function $M_{j,u}$. As in Refs. \cite{Tasca18a,paul18}, we choose the
bin functions to be periodic square waves
\begin{equation}
M_{j,u}\left(z;T\right)=\begin{cases}
1, & u s_j\leq z\;\left(\textrm{mod }T_j\right)\leq\left(u+1\right)s_j\\
0, & \textrm{otherwise}
\end{cases},\label{eq:mask-definition}
\end{equation}
specified by
the period $T_j$ and bin width $s_j=T_j/d$, so that $d$ can indeed be considered
as a dimensionality parameter. In Fig.
\ref{fig:vecs} c) we illustrate the periodic bin
function for the particular case of $d=3$.
The outcome
probabilities produced by the set of projectors \eqref{Projectors}
then define the PCG of the probability distribution associated with the phase-space
variable $q_{j}$. Since we work with dimensionless variables,
the parameters $T_j$ and $s_j$ are also  dimensionless.  
\par
Without loss of generality, we assume that $\theta_0=0$ and order the variables so that $\theta_j>\theta_k$ whenever $j>k$. The condition for mutual unbiasedness  of these PCG operators 
was obtained in Ref. \cite{paul18}  as a linear relation between the dimensionality parameter 
and the product of periodicities of a pair of PCG operators:
\begin{equation}
T_j T_k m_{jk} = 2 \pi d |\sin \theta_{jk}|,
\label{eq:MUMcond1}
\end{equation}
where $\theta_{jk} \equiv \theta_j - \theta_k$ as before, and $m_{jk}$ is a positive integer.  The condition \eqref{eq:MUMcond1} is supplemented by the requirement that all $m_{jk}$ are coprime with $d$, i.e.
\begin{equation}
\frac{m_{jk} n}{d} \notin \mathbb{N},\quad\,\, \forall \, n = 1,\dots, d-1.
\label{eq:MUMcond1-b}
\end{equation}
\par
\section{Results}

\subsection{Maximal number of PCG mutually unbiased measurements}

The main technical aim of this paper is to study the dependence between the maximum allowed value of $R$, such that all PCG measurements are mutually unbiased according to the operational definition (\ref{MUB1}), and the dimensionality $d$. We remind the reader that for discrete systems (instead of PCG) we know that $R\leq d+1$ (and the bound is tight for all $d$ that is a power of a prime number), while for continuous systems we have $R\leq 3$. Therefore, the way in which dimensionality bounds the number of mutually unbiased measurements is a signature of discrete/continuous type of behavior. Below, we prove that for PCG:
\begin{equation}
\begin{cases}
R\leq p + 1  & \qquad\textrm{general }d \\
R\leq d+1 & \qquad\textrm{prime }d\\
R\leq3 & \qquad\textrm{even } d
\end{cases},
\end{equation}
where
$p$ is the smallest prime factor of $d$.  For composite (non-prime) dimensions, the fact that $p < d$ shows that the case of  prime dimension is maximal, similarly to the discrete scenario. On the other hand, when $d$ is a prime power, i.e. $d=p^k$, here we still have $R \leq p+1$, so that the pattern known from discrete systems \cite{Ivonovic1981,wootters1989,Bandyopadhyay2002,klappenecker2003} is not reproduced.  All distinct cases described above are collected in Table \ref{Tableb1}.

\begin{table}[t!]
\begin{centering}
\begin{tabular}{|c|c|}
\hline 
Dimension $d$ & Pattern for maximal $R$ \tabularnewline 
\hline 
\hline 
general &  sub-discrete \tabularnewline
\hline 
even  & continuous\tabularnewline
\hline 

odd, prime & discrete \tabularnewline
\hline 
odd, prime powers & different\tabularnewline
\hline 
\end{tabular}\caption{Different types of behavior for maximal number $R$ of PCG MUMs.\label{Tableb1}}
\par\end{centering}
\end{table}

\textbf{Proof.} Let us assume that  $\sin \theta_j \geq 0$, which defines variables $q_j$ in the upper semi-plane of the phase space.  This is not a restriction, as variables in the lower half-plane can be taken to the upper half-plane by a reflection through the origin: $q_j \rightarrow -q_j$.  Using the assumption that $\theta_0=0$, the $R-1$ distinct conditions stemming from Eq. \eqref{eq:MUMcond1} with $k=0$ can be rewritten to the form:
\begin{equation}
T_j  = \frac{2 \pi d \sin \theta_{j}}{m_{j0} T_0}.
\label{eq:periods}
\end{equation}
These are solutions for all periods except $T_0$.  
Plugging the above equations into \eqref{eq:MUMcond1} with $j>k > 0$, we have
 \begin{equation}
T_j T_k m_{jk} = 2 \pi d |\sin \theta_{jk}| = \frac{ (2 \pi d)^2 \sin \theta_{j}\sin \theta_{k}m_{jk}} {T_0^2 m_{j0}m_{k0}}. 
\label{eq:MUMcond2}
\end{equation}
This can be recast as 
  \begin{equation}
 \frac{m_{jk}} {m_{j0}m_{k0}} = \frac{T_0^2}{2 \pi d} (\cot \theta_k - \cot\theta_j).
\label{eq:MUMcond4}
\end{equation}
There is no absolute value anymore since $\cot \theta_k > \cot\theta_j$ due to the monotonic behavior of the cotangent and the assumed ordering and range of the angles. The condition Eq. \eqref{eq:MUMcond4} for $k=1$ and $j\geq2$, further gives
\begin{equation}\label{angle}
\cot\theta_{j}=\cot\theta_{1}-\frac{2\pi d}{T_{0}^{2}}\frac{m_{j1}}{m_{j0}m_{10}}.
\end{equation}
These are the solutions for all the angles, except the first non-trivial one. They can now be substituted into the remaining relations
for $j>k>1$, producing the following set of constraints involving only natural numbers:
\begin{equation}
m_{j1}m_{k0}-m_{j0}m_{k1}=m_{jk}m_{10}.\label{eq:ms}
\end{equation}
\par
Let us write $d=p p^\prime$, where $p$ is the smallest non-trivial prime factor of $d$, so that $p=d$ and $p^\prime =1$ if $d$ is prime. Observe that condition \eqref{eq:MUMcond1-b} implies that all $m_{jk}$ are coprime with $p$, as can be concluded by taking the particular instance $n=p^\prime$. This observation has two consequences. First, all  $m_{jk}$ possess a modular multiplicative inverse $\tilde m_{jk}$ (with respect to $p$) which additionally is an integer. Formally, we say that $m_{jk}\tilde m_{jk}\equiv 1\; (\text{mod }p)$.  Second, the product $m_{jk}m_{10}$ is also coprime with $p$, fact that is mathematically expressed as $m_{jk}m_{10}\not\equiv 0\; (\text{mod }p)$.   
Therefore, due to (\ref{eq:ms}), we obtain the constraint
\begin{equation}
m_{j1}m_{k0}\not\equiv m_{j0}m_{k1}\; (\text{mod }p)\qquad \forall \,j>k>1.\label{eq:ms2p}
\end{equation}

Let us introduce auxiliary parameters $\chi_k=m_{k0}\tilde m_{k1}$, for $k=2,\ldots,R-1$.  After ``modulo $p$'' multiplication by the integer $\tilde m_{k1}\tilde m_{j1}$, Eq. (\ref{eq:ms2p}) simplifies to the form
\begin{equation}
\forall_{j>k>1}\quad\chi_k\not\equiv \chi_j\; (\text{mod }p).
\end{equation}
To satisfy the above condition, each $\chi_k$ must belong to a different \textit{congruence class modulo $p$}. Therefore, at most we can have as many different values of $\chi_k$, as is given by the number of congruence classes, which is equal to $p-1$ since $p$ is a prime number. The final result, i.e. $R\leq p +1$ follows, since apart from $p$ distinct values of $\chi_k$ we need to include the phase-space variables $q_0$ and $q_1$.  If $d$ is prime, we remember that we have $p=d$, giving $R \leq d+1$. 

In the case of even $d$, the above result immediately implies that there are at most $R=3$ mutually unbiased PCG observables because $p=2$ for all even dimensions.

\subsection{Symmetric configuration}
There is  quite a bit of freedom in the constraint \eqref{eq:MUMcond4} concerning the period $T_0$, as well as the angles $\theta_j$. Elaborating on the solution (\ref{angle}), with some specification, we can however construct a useful, symmetric recipe for mutually unbiased measurements. Let us choose angles that are distributed at integer multiples of a fixed angle $\theta$, such that $\theta \leq \pi/R$ and $\theta_j= j \theta$. Moreover, we will choose $T_0 =\sqrt{\pi d  \tan \theta}$, so that Eq.  (\ref{angle}) divided by $\cot \theta$ assumes a simplified form
\begin{equation}\label{angle2}
\frac{\cot j\theta}{\cot\theta}=1-2\frac{m_{j1}}{m_{j0}m_{10}}.
\end{equation}
The right hand side of this condition is, by construction, a rational number. Since
\begin{equation}\label{Cotj}
\frac{\cot j\theta}{\cot\theta}=\frac{\sum_{l\,\textrm{even}}\left(-1\right)^{\frac{l}{2}}\left(\begin{array}{c}
j\\
l
\end{array}\right)\tan^{l}\theta}{\sum_{l\,\textrm{odd}}\left(-1\right)^{\frac{l-1}{2}}\left(\begin{array}{c}
j\\
l
\end{array}\right)\tan^{l-1}\theta},
\end{equation}
we can see that the left hand side is rational as well, whenever
\begin{equation}
\tan \theta = \sqrt{Q},
\label{eq:MUMcond6}
\end{equation}
with $Q$ being a nonnegative rational number. This is because both numerator and denominator in Eq. \eqref{Cotj} only contain even powers of $\tan \theta$ which are equal to powers of $Q$. This construction does further allow us to adjust all $m_{j0}$ and $m_{j1}$. 
\par
\subsection{Experiment}
To confirm and explore our results we performed an experiment using the transverse spatial degrees of freedom of single photons produced by an attenuated laser source, and detected by a single photon detector. 
The transverse spatial degrees of freedom of a paraxial optical field are analogous to position and momentum variables in quantum mechanics \cite{marcuse82}.  Here the near-field  and far-field variables with respect to a transverse reference plane play the role of position and momentum, respectively. One can perform arbitrary rotations in the phase space formed by these variables using fractional Fourier transforms (FrFTs) \cite{tasca11}. In this way, we can transform the position to a variable $q_j$ and then  an arbitrary $q_k$ by a sequence of two FrFTs parametrized by the angles $\theta_j$ and $\theta_{kj}=\theta_k-\theta_j$, respectively \cite{ozaktas01}.   The projective measurements (rank $> 1$) were realized using amplitude masks to select only the appropriate values of $q_j$ for each projection, following the definition of the bin functions given in Eq. \eqref{eq:mask-definition} and illustrated in Fig. \ref{fig:vecs}.
\begin{figure}[th]
\includegraphics[width=8cm]{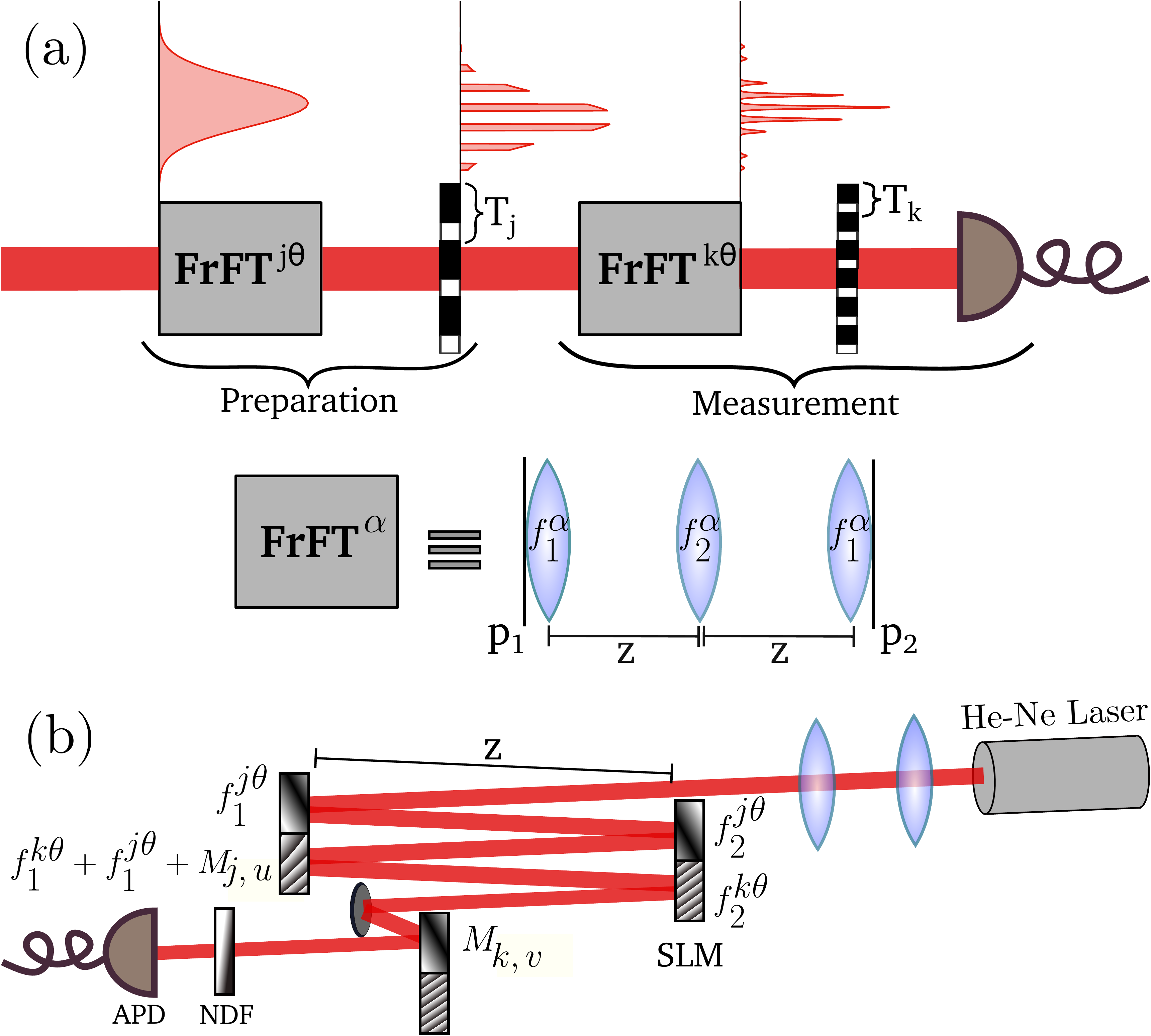}
\caption{ a) Schematic of the experiment.  Fractional Fourier Transforms (FrFT) and periodic amplitude masks (M) are used to prepare and measure the transverse spatial profile of the laser beam.  All output light is incident on full-field single-photon detector.  b) Experimental setup using spatial light modulators (SLM) for preparation and measurement. The output of a $632.8$nm He-Ne laser is enlarged and collimated using two lenses. The FrFTs are performed modulating the phase accordingly to the quadratic phase of lenses with the right focal distances. The use of SLMs allows us to synthesize any order FrFT, which would be challenging with actual lenses. The amplitude masks are also implemented using phase-only modulators applying diffraction gratings and collecting only the first diffraction order. Finally, the beam is attenuated with a neutral density filter (NDF) and detected by a single photon detector. }
\label{fig:exp}
\end{figure}
\par
Optical FrFTs and amplitude masks were used to both prepare eigenstates with respect to one PCG measurement, and perform different PCG measurements on them, as shown schematically in Fig. \ref{fig:exp}.  Both the FrFTs as well as the amplitude masks were implemented using spatial light modulators (SLMs), as described in App. \ref{exp}.  In the preparation stage, a FrFT of order $\theta_j$ was implemented on the transverse profile, followed by the application of an amplitude mask $M_{j,u}$ of period $T_j$. This corresponded to the preparation of the initial state $u$, in basis $j$. The measurement stage consisted of an FrFT of order $\theta_k$ and an amplitude mask $M_{k,v}$, projecting onto outcome $v$ in basis $k$. The full field of the resulting output beam was  detected with a single photon detector, as depicted in Fig. \ref{fig:exp}-(a). In principle, the use of this FrFT scheme with SLMs allows us to perform any combination of preparation and measurement in a black-box like manner, giving as inputs the dimension $d$, the preparation  index $u\in \{0,\dots,d-1\}$, the phase space directions $\theta_j$ and $\theta_k$,  and the respective mask periods $T_j$ and $T_k$, and obtaining as output the probability of each measurement outcome $v\in \{0,\dots,d-1\}$.
 \par
We tested the case of $R=4$ MUMs in the symmetric configuration with dimension parameter $d=3$ and $\theta=\pi/4$, so that $\tan \theta=1$ fulfills relation (\ref{eq:MUMcond6}), for all sixteen combinations of preparation and measurement.   We chose the period of the $0$th mask to be $T_0'=93$ pixels so that with the choice $m_{10}=m_{30}=1$ we have $T_1'=T_3' = 132$ pixels, and for $m_{20}=2$ we have  $T^\prime_2 = 93$ pixels.  These periods are the closest values in integer number of pixels to those that  satisfy all conditions \eqref{eq:MUMcond4} for these $m_{jk}$ (see App. \ref{exp}).
\par
We tested MUM conditions between preparations $j=0,1,2,3$ and measurements $k=0,1,2,3$. 
For each measurement $k$, the detection mask was scanned in all three positions ($v=0,1,2$), and the number of photo-counts 
$C_k^{(v)}$ registered.   We then calculated the detection probabilities $p_k^{(v)}=C_k^{(v)}/\sum_v C_k^{(v)}$. This was repeated for each pair of preparation/measurement observables $j/k$.   
  \par
  To evaluate the mutual unbiasedness of the measurement results, we calculated the Shannon entropy $E=-\sum_v p_k^{(v)} \log_2 p_k^{(v)}$ of the probability distributions (see App. \ref{probs}).   The results are shown in Table \ref{tab:1}.  When the preparation basis $j$ is different from the measurement basis $k$, we obtain entropies that are close to the maximum $\log_2 3 \approx 1.5849$, indicating that the probability distributions are nearly uniform.  
To demonstrate that this is indeed due to mutual unbiasedness and not simply a result of some source of homogeneous noise, we tested the setup by realizing preparation and measurement in the same phase-space direction.  In this case, we expect a deterministic probability distribution given by 
$p_k^{(v)}=\delta_{v,u}$, where $u$ is the preparation state.  In all cases we observed a single probability $p_k^{(v=u)} > 0.983(5)$ and the other two 
$p_k^{(v \neq u)} <  0.0131(4)$. With a deterministic probability distribution, one expects a null value for the entropy.  However, the overall $2\%$ background noise is enough to result in the non-null entropy values shown in the diagonal of Table \ref{tab:1}.  Still, this is sufficient to verify that the setup is functioning properly.  Our results thus confirm mutual unbiasedness for $R=d+1=4$ PCG measurements with properly chosen periods. 
\begin{table}
\begin{tabular}{c|c|c|c|c|c|}
\multicolumn{1}{c}{\multirow{6}{*}{\begin{turn}{90}
\hspace{-0.5cm} Preparation
\end{turn}}} & \multicolumn{5}{c}{Measurement}\tabularnewline
\cline{2-6} 
 &  & 0 & 1 & 2 & 3\tabularnewline
\cline{2-6} 
 & 0 & $0.161(3)$ & $1.5846(2)$ & $1.579(1)$ & $1.5847(2)$\tabularnewline
\cline{2-6} 
 & 1 & $1.5841(8)$ & $0.143(3)$ & $1.583(3)$ & $1.584(4)$\tabularnewline
\cline{2-6} 
 & 2 & $1.5846(1)$ & $1.5848(1)$ & $0.140(5)$ & $1.5838(1)$\tabularnewline
\cline{2-6} 
 & 3 & $1.5844(2)$ & $1.5847(1)$ & $1.5844(5)$ & $0.162(3)$\tabularnewline
\cline{2-6} 
\end{tabular}
\caption{Entropy value for different preparation and measurement directions in the case $R=4$ and $d=3$ with periodicities satisfying all the MU conditions \eqref{eq:MUMcond1}. }
\label{tab:1}
\end{table}

\par
\begin{figure}[t!]
\includegraphics[width=8cm]{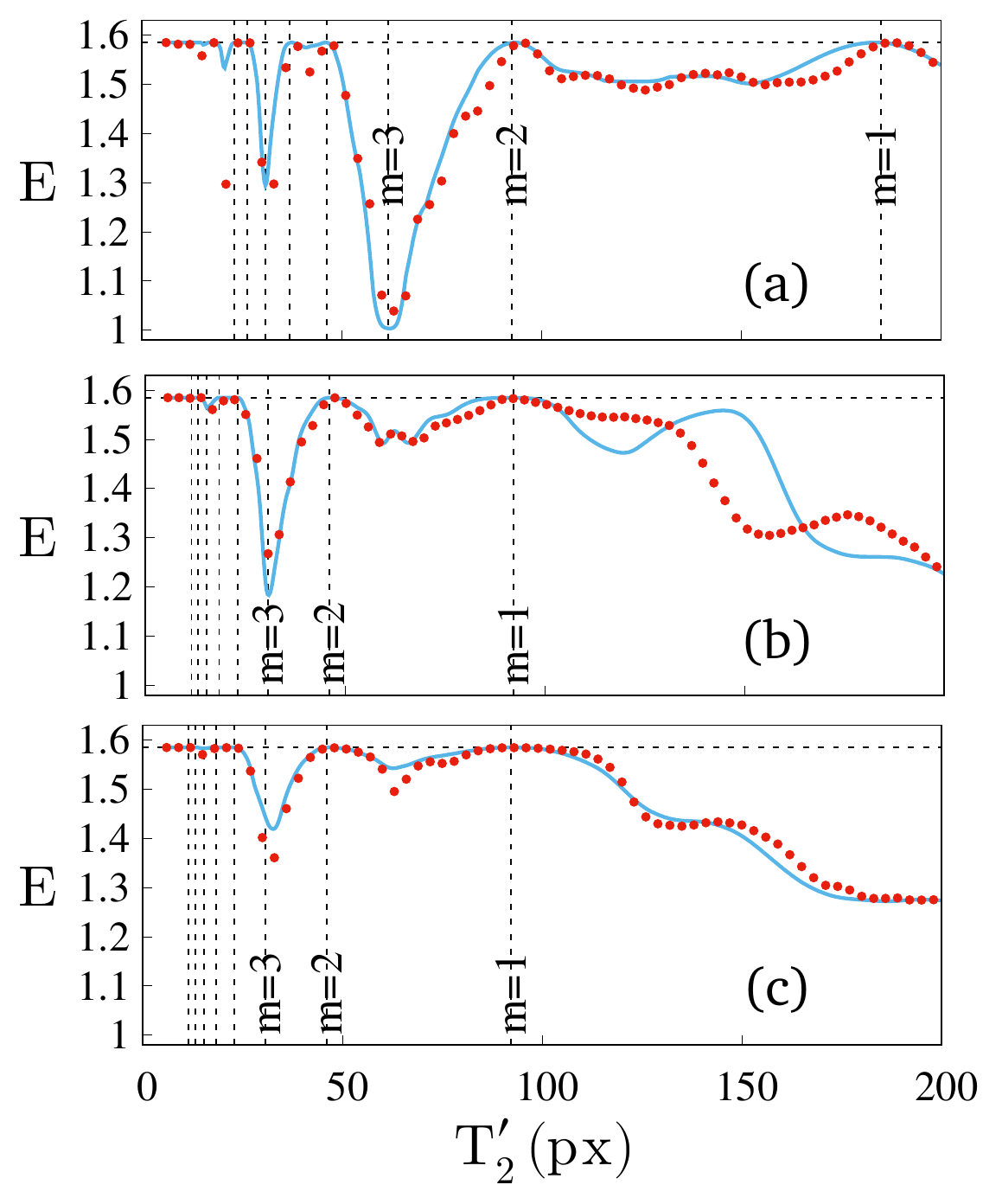}
\caption{{Experimental results for the entropy for $R=4$ PCG measurements with a dimension parameter $d=3$, using angle $\theta=\pi/4$. Entropy is plotted as a function of the physical period $T^{\prime}_2$ in units of pixels (px) of the measurement mask, with preparation fixed at position ``0" and  period that satisfies the mutual unbiasedness condition \eqref{eq:periods} for $m_{j0}=1$.  The measurement direction is fixed at $k=2$ and the entropy is determined for preparation directions: a) $j=0$, b) $j=1$, and c) $j=3$.  The blue curves correspond to theoretical predictions for the experimental initial state. The red dots represent the experimental data, the error bars are smaller than the dots. The vertical dashed lines indicate the special values of period satisfying \eqref{eq:MUMcond1} for different values of $m_{j2}$. }} 
\label{fig:results}
\end{figure}

  To further explore mutual unbiasedness in four phase space directions and validate our results, we tested the mutual unbiasedness conditions between preparations $j=0,1,3$ and measurement $k=2$ as a function of the period $T^\prime_2$ used in the measurement stage, obtaining experimental results shown in Fig. \ref{fig:results}. {The theoretical prediction (solid curve in the figure) is the result of integrating the theoretical probability density (such as the ones shown in Fig \ref{fig:estados}) over the appropriate bin regions.}  We can see that at several places the entropy reaches its maximum value of $\log_2 3 \approx 1.5849$, which indicates that the probability distribution is uniform, corresponding to mutual unbiasedness.  Vertical lines show values at which the period $T^\prime_2$ corresponds to allowable $m_{j2}$ values.  In all plots, we can observe that the entropy decreases greatly when $m_{j2}=3$, which is not allowed by Eq. \eqref{eq:MUMcond1-b} when $d=3$.  Furthermore, the MUM condition is reached whenever the period $T^\prime_2$ equals $93\,$pixels, as  expected from the derived requirements. 
  Moreover, it corresponds to $m_{21}=m_{32}=1$ and $m_{20}=2$, as predicted by our theoretical results and previously confirmed experimentally in Table \ref{tab:1}.  Similar results were obtained for all combinations of preparation and measurement. The departure of the experimental points from the theoretical curve for larger periods (smaller bins) can be intuitively understood by considering that when the period is small, unaccounted propagation errors such as diffraction affect a large number of the bins more or less equally, so that the errors are homogeneous and do not significantly affect the output probability distribution, which is already close to uniform.  However, as the period increases, the regions significantly affected by the experimental errors can become concentrated in only a few bins, which begins to modify the final probability distribution more significantly.
\par
Let us briefly discuss experimental errors.
In Fig.  \ref{fig:results} we observe a limiting factor of our experimental implementation, and of the PCG construction in general.  That is, periods $T_j$ satisfying condition \eqref{eq:periods} are inversely proportional to $m_{j0}$, and thus become closer together as $m_{j0}$ increases (small $T$), eventually reaching the limits imposed by the spatial resolution of our optical system.  In addition to causing experimental errors, this loss of resolution sets a limit for accessible values of $m_{j0}$. In our system this occurs at around $m_{20} \sim 14$, for which $T^\prime_2(m_{20}=13)- T^\prime_2(m_{20}=14) \approx 1$pixel. 
\par
On the other hand, when the period is large, we essentially recover the case of standard coarse graining, where only the central set of bins have non-zero probability.  Thus, the interesting properties that arise from the periodicity of the bin functions is lost.  In this case, additional experimental errors such as the alignment of the beam becomes more relevant, since with misalignment we lose mutually unbiasedness.  This is not as relevant when the period is small, since the field covers many bins.    Our error bars take into account only the uncertainty resulting from Poissonian count statistics, and do not consider any other source of experimental error, such as misalignment (transverse and angular displacements).   To avoid these errors it is best to work in the middle region, where the periods are neither too large nor too small.  This gives a robustness to alignment errors, as well avoids any possible resolution issues with the spatial light modulator.  We expect that other physical systems will encounter similar types of constraints.

\section{Discussions}  We have shown that it is possible to define a general set of $R$ mutually unbiased observables on a continuous variable system that produces a discrete set of $d$ outcomes, while at the same time displaying behavior that is neither discrete nor continuous.  This conclusion is drawn by evaluating the behavior of these measurements under conditions for mutual unbiasedness.   In particular, for prime $d$ we have shown that there can be $d+1$ mutually unbiased observables, which is reminiscent of the discrete behavior.  However, for composite $d$, we can have at most $p+1$, where $p$ is the smallest non-trivial prime factor of $d$. Thus, for even $d$, we can define at most 3 mutually unbiased measurements, which is similar to the continuous case. However, when $d=p^k$ is the power of a prime number, we have $p+1$,  which is distinct from both the discrete and continuous cases.  
\par
An interesting aspect of this scenario is that it allows us to determine the non-trivial maximum number of mutually unbiased measurements for all values of $d$, which has not yet been achievable for discrete systems.  
We have corroborated and explored the practicality of our theoretical result with an experiment observing mutual unbiasedness for 4 measurements with $d=3$ outcomes using the transverse spatial degrees of freedom of photons.  This is already beyond what can be achieved  in the continuous regime.  This demonstration suggests that it should be interesting to explore these types of PCG measurements in the detection and utilization of spatially entangled photon pairs outside of the usual near-field/far-field scenario \cite{tasca08,tasca09a,Paul16}.  Moreover, we expect that our results will find utility in applications that exploit unbiasedness, such as randomness generation, for example. 

\begin{acknowledgments}
 The authors thank R. V. Nery, E. C. Paul, A. Z. Khoury, P. H. Souto Ribeiro, R. L. de Matos Filho and G. Lima  for insightful discussions. T.L.S., D.S.T and S.P.W. acknowledge partial financial support from the Brazilian agencies CNPq (304196/2018-5, 431804/2018-4), FAPERJ
(PDR10 E-26/202.802/2016,  E-26/202.7890/2017), CAPES
(PROCAD2013) and the INCT-IQ (465469/2014-0). S.P.W acknowledges support from the chilean Fondo Nacional de Desarrollo Cient\'{i}fico y Tecnol\'{o}gico (FONDECYT) (1200266) and  ANID – Millennium Science Initiative Program – ICN17\_012. \L.R. acknowledges funding by the Foundation for Polish Science (IRAP project, ICTQT, Contract No. 2018/MAB/5, cofinanced by EU within Smart Growth Operational Programme).
\end{acknowledgments}
\vspace{10pt}

\appendix

\section{\uppercase{Experimental Implementation of Mutually Unbiased Measurements}}
\label{exp}

\begin{figure}[b!]
\includegraphics[width=6cm]{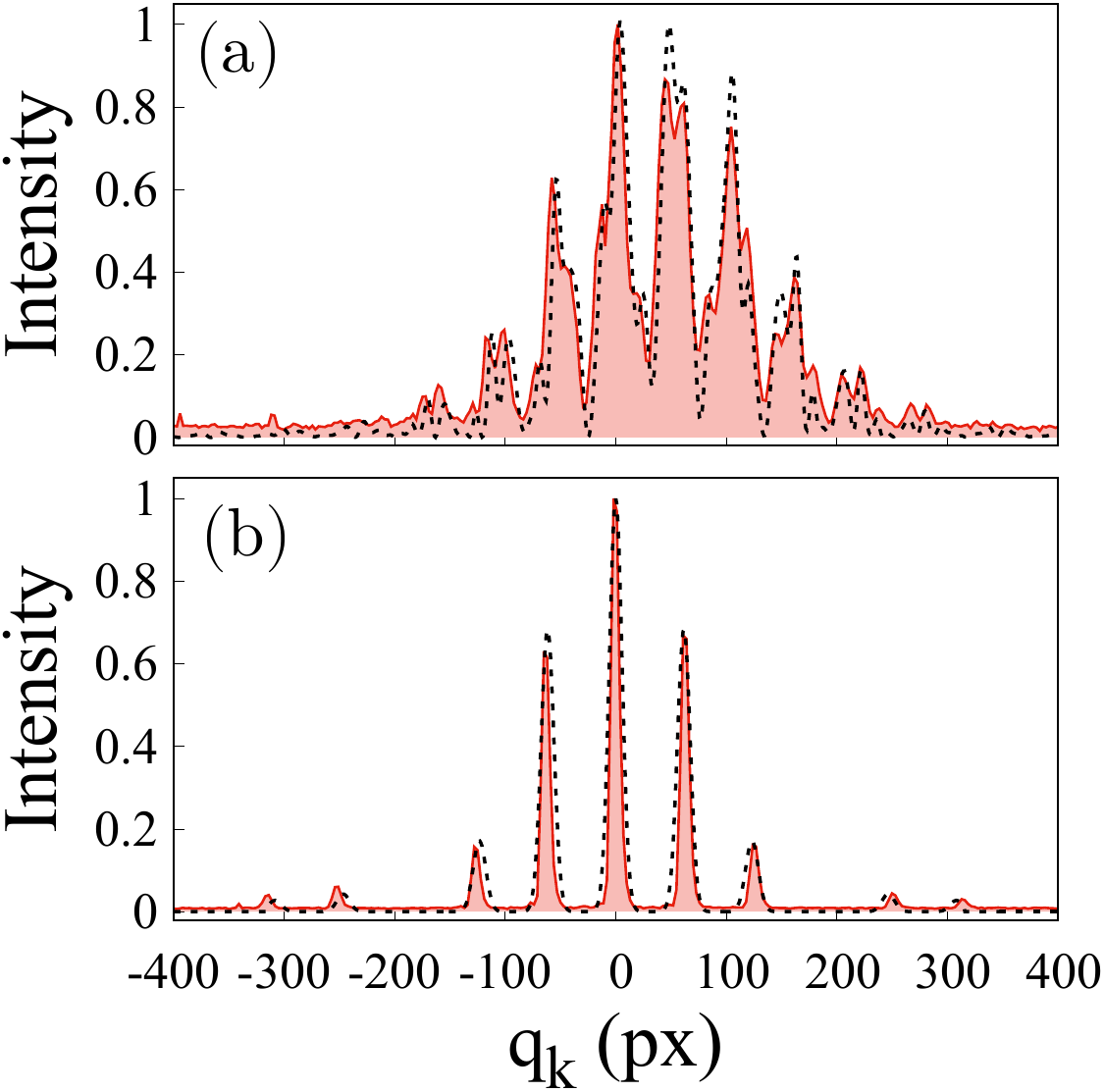}
\caption{{Experimental intensity profiles (red solid curves) at the output SLM obtained by scanning the SLM with a narrow slit aperture while measuring the number of photo-counts. The intensity profile is proportional to the probability density of $q_k$. The localized state prepared in direction $j=0$ is represented in directions (a) $k=1$ and (b) $k=2$. The experimental result is compared to the theoretical curves obtained by numerically propagating the initial state (dashed black curves), i.e. by applying the periodic masks and FrFTs numerically onto a Gaussian wave function representing the initial profile of the optical field.}}
\label{fig:estados}
\end{figure}

The initial state in position is fixed and is prepared as a collimated Gaussian beam with beam radius $(2.54\pm0.06)\,$mm at the plane of the first SLM.  The beam was prepared and measured with rotations in phase space, performed with FrFTs, and amplitude masks.  To realize the FrFTs, we use the scheme described in detail in Ref. \cite{Rodrigo09}, where SLMs are used to imitate three lenses by imprinting quadratic phases on the optical field (see \ref{fig:exp}).  
{ The lenses are separated by a distance $z$. If the focal distances satisfy $f_{1}^\alpha=z\left(1-\frac{\cot(\alpha/2)}{2}\right)^{-1}$ and $f_2^\alpha=z(1-\sin\alpha)^{-1}$ then the transverse profile of the optical field in the plane $p_2$ is the FrFT of order $\alpha$ of the field in the plane $p_1$ up  to a scaling factor that is independent of the FrFT order}.
The SLMs were also used to imprint amplitude masks with physical period $T'_j$, which define the discrete bins in our measurements. 
Using the three-lens FrFT scheme introduces a scaling factor independent of the rotation angle, such that dimensionless ($T_j$) and physical ($T'_j$) periods are related by $T'_j=\sqrt{\frac{\lambda z}{\pi}}T_j$, where $z=0.29\,$m is the distance between the lenses and $\lambda=632.9\,$nm is the He-Ne laser wavelength. Moreover, for practical reasons, the physical periods are given in units of SLM pixels. The pixel size of the Holoeye SLMs used here is $8\,\mu\text{m}$.  We chose the period of the $0$th mask to be $T_0'=93$ pixels, since this value is the closest integer number to the value of the exact {solution} ($92.7476$ pixels).  Using \eqref{eq:periods}, and choosing $m_{10}=m_{30}=1$ we have $T_1'=T_3' = 131.165$ pixels, approximated by $132$ pixels so the bin width is an integer number of pixels.   For $m_{20}=2$ we have  $T^\prime_2 = 92.7476 \approx 93$ pixels.  These values of $m_{jk}$ satisfy all conditions \eqref{eq:MUMcond4}.

To test the experimental setup, we scanned the output SLM with  a narrow slit aperture so that we could measure the final transverse intensity profile of light produced by the optical transformations acting on the initial state (all but final amplitude mask $M_{k,v}$).  Some results are exemplified in Fig. \ref{fig:estados}. This intensity profile should be proportional to the probability density obtained from the theoretical initial state propagated through the preparation mask and two FrFTs. 
In this way, Fig. \ref{fig:estados} reveals the qualitative agreement between our theoretical description and the optical setup. The theoretical probability densities obtained as described are used to predict the results when the PCG measurements are performed.  

\section{\uppercase{Experimental probability distributions}\label{probs}}

\begin{table}[ht]
\begin{tabular}{c|c|c|c|c|c|}
\multicolumn{1}{c}{\multirow{6}{*}{\begin{turn}{90}
\hspace{-0.5cm} Preparation
\end{turn}}} & \multicolumn{5}{c}{Measurement}\tabularnewline
\cline{2-6} 
 &  & 0 & 1 & 2 & 3\tabularnewline
\cline{2-6} 
 & 0 & $-$ & $0.65(2)$ & $1.3(3)$ & $0.4(2)$\tabularnewline
\cline{2-6} 
 & 1 & $0.7(3)$ & $-$ & $1.1(3)$ & $0.2(1)$\tabularnewline
\cline{2-6} 
 & 2 & $0.13(5)$ & $0.70(9)$ & $-$ & $0.45(6)$\tabularnewline
\cline{2-6} 
 & 3 & $0.40(8)$ & $0.2(6)$ & $0.73(5)$ & $-$\tabularnewline
\cline{2-6} 
\end{tabular}
\caption{Kullback-Leibler divergence ($10^{-3}$) from the uniform distribution of the probability distributions of Fig. \ref{fig:probs}. }
\label{tab:2}
\end{table}

\begin{figure}[!hb]
\includegraphics[width=0.99\columnwidth]{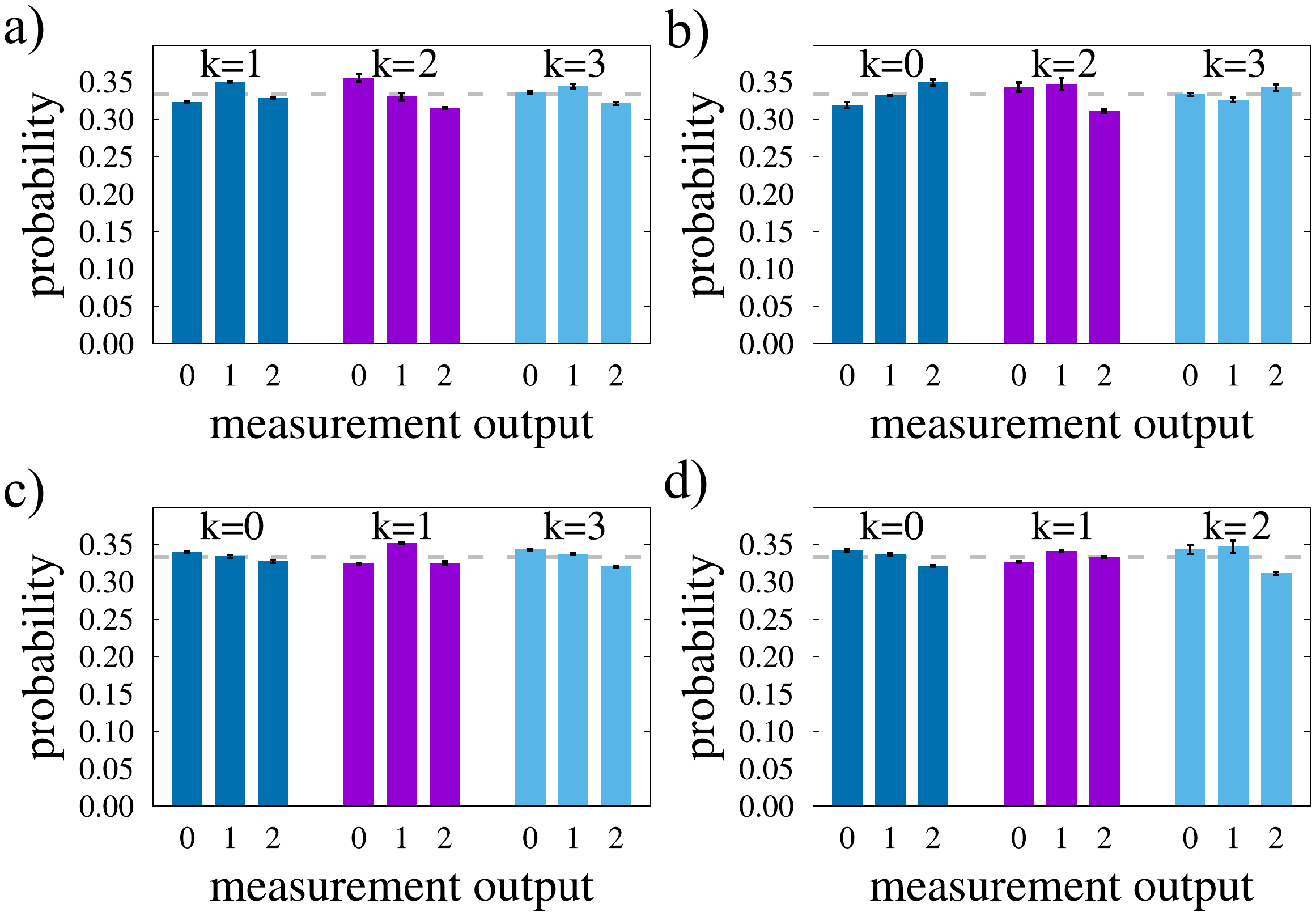}
\caption{Experimental outcome probabilities used to calculate the Shannon entropies  shown in Table \ref{tab:1} for preparation directions a) $j=0$, b) $j=1$, c) $j=2$, and d) $j=3$. Only the cases when the measurement direction $k$ is different from that used to prepare the localized state are presented. The dashed horizontal line corresponds to $p=\frac{1}{d}=\frac{1}{3}$.  }
\label{fig:probs}
\end{figure}

For the case when  the four PCG measurements satisfy the pairwise MUM condition, the values of the Shannon entropy shown in Table \ref{tab:1} are close to the maximum of $\log(d)$ suggesting that in that case we obtained flat probability distributions or unbiasedness. Fig. \ref{fig:probs} show the probability distributions obtained in our experiment. For all combinations $j,k$ of preparation and measurement directions we observe a nearly flat distribution. It can be confirmed further by calculating the Kullback-Leibler divergence from the ideal uniform distribution with
\begin{equation}
 D(P||Q)=\sum_{v=0}^{d-1} p^{(v)}\log\left(\frac{p^{(v)}}{q^{(v)}}\right),
\end{equation}
where $p^{(v)}$ is the experimental probability of outcome $v$ and $q^{(v)}=\frac{1}{d}$ is the uniform distribution. The values of the Kullback-Leibler divergence for the unbiased PCG configuration is below $0.0013(3)$ as can be seen in Table \ref{tab:2}.

\bibliography{bibliographyDMUMPCG}

\begin{thebibliography}{40}%
\makeatletter
\providecommand \@ifxundefined [1]{%
 \@ifx{#1\undefined}
}%
\providecommand \@ifnum [1]{%
 \ifnum #1\expandafter \@firstoftwo
 \else \expandafter \@secondoftwo
 \fi
}%
\providecommand \@ifx [1]{%
 \ifx #1\expandafter \@firstoftwo
 \else \expandafter \@secondoftwo
 \fi
}%
\providecommand \natexlab [1]{#1}%
\providecommand \enquote  [1]{``#1''}%
\providecommand \bibnamefont  [1]{#1}%
\providecommand \bibfnamefont [1]{#1}%
\providecommand \citenamefont [1]{#1}%
\providecommand \href@noop [0]{\@secondoftwo}%
\providecommand \href [0]{\begingroup \@sanitize@url \@href}%
\providecommand \@href[1]{\@@startlink{#1}\@@href}%
\providecommand \@@href[1]{\endgroup#1\@@endlink}%
\providecommand \@sanitize@url [0]{\catcode `\\12\catcode `\$12\catcode
  `\&12\catcode `\#12\catcode `\^12\catcode `\_12\catcode `\%12\relax}%
\providecommand \@@startlink[1]{}%
\providecommand \@@endlink[0]{}%
\providecommand \url  [0]{\begingroup\@sanitize@url \@url }%
\providecommand \@url [1]{\endgroup\@href {#1}{\urlprefix }}%
\providecommand \urlprefix  [0]{URL }%
\providecommand \Eprint [0]{\href }%
\providecommand \doibase [0]{https://doi.org/}%
\providecommand \selectlanguage [0]{\@gobble}%
\providecommand \bibinfo  [0]{\@secondoftwo}%
\providecommand \bibfield  [0]{\@secondoftwo}%
\providecommand \translation [1]{[#1]}%
\providecommand \BibitemOpen [0]{}%
\providecommand \bibitemStop [0]{}%
\providecommand \bibitemNoStop [0]{.\EOS\space}%
\providecommand \EOS [0]{\spacefactor3000\relax}%
\providecommand \BibitemShut  [1]{\csname bibitem#1\endcsname}%
\let\auto@bib@innerbib\@empty
\bibitem [{\citenamefont {Revzen}\ \emph {et~al.}(2005)\citenamefont {Revzen},
  \citenamefont {Mello}, \citenamefont {Mann},\ and\ \citenamefont
  {Johansen}}]{revzen05}%
  \BibitemOpen
  \bibfield  {author} {\bibinfo {author} {\bibfnamefont {M.}~\bibnamefont
  {Revzen}}, \bibinfo {author} {\bibfnamefont {P.~A.}\ \bibnamefont {Mello}},
  \bibinfo {author} {\bibfnamefont {A.}~\bibnamefont {Mann}},\ and\ \bibinfo
  {author} {\bibfnamefont {L.~M.}\ \bibnamefont {Johansen}},\ }\bibfield
  {title} {\bibinfo {title} {Bell's inequality violation with non-negative
  wigner functions},\ }\href {https://doi.org/10.1103/PhysRevA.71.022103}
  {\bibfield  {journal} {\bibinfo  {journal} {Phys. Rev. A}\ }\textbf {\bibinfo
  {volume} {71}},\ \bibinfo {pages} {022103} (\bibinfo {year}
  {2005})}\BibitemShut {NoStop}%
\bibitem [{\citenamefont {Gilchrist}\ \emph {et~al.}(1998)\citenamefont
  {Gilchrist}, \citenamefont {Deuar},\ and\ \citenamefont
  {Reid}}]{gilchrist98}%
  \BibitemOpen
  \bibfield  {author} {\bibinfo {author} {\bibfnamefont {A.}~\bibnamefont
  {Gilchrist}}, \bibinfo {author} {\bibfnamefont {P.}~\bibnamefont {Deuar}},\
  and\ \bibinfo {author} {\bibfnamefont {M.~D.}\ \bibnamefont {Reid}},\
  }\bibfield  {title} {\bibinfo {title} {Contradiction of quantum mechanics
  with local hidden variables for quadrature phase amplitude measurements},\
  }\href {https://doi.org/10.1103/PhysRevLett.80.3169} {\bibfield  {journal}
  {\bibinfo  {journal} {Phys. Rev. Lett.}\ }\textbf {\bibinfo {volume} {80}},\
  \bibinfo {pages} {3169} (\bibinfo {year} {1998})}\BibitemShut {NoStop}%
\bibitem [{\citenamefont {Banaszek}\ and\ \citenamefont
  {W\'odkiewicz}(1998)}]{banaszek98}%
  \BibitemOpen
  \bibfield  {author} {\bibinfo {author} {\bibfnamefont {K.}~\bibnamefont
  {Banaszek}}\ and\ \bibinfo {author} {\bibfnamefont {K.}~\bibnamefont
  {W\'odkiewicz}},\ }\bibfield  {title} {\bibinfo {title} {Nonlocality of the
  einstein-podolsky-rosen state in the wigner representation},\ }\href
  {https://doi.org/10.1103/PhysRevA.58.4345} {\bibfield  {journal} {\bibinfo
  {journal} {Phys. Rev. A}\ }\textbf {\bibinfo {volume} {58}},\ \bibinfo
  {pages} {4345} (\bibinfo {year} {1998})}\BibitemShut {NoStop}%
\bibitem [{\citenamefont {Banaszek}\ and\ \citenamefont
  {W\'odkiewicz}(1999)}]{banaszek99}%
  \BibitemOpen
  \bibfield  {author} {\bibinfo {author} {\bibfnamefont {K.}~\bibnamefont
  {Banaszek}}\ and\ \bibinfo {author} {\bibfnamefont {K.}~\bibnamefont
  {W\'odkiewicz}},\ }\bibfield  {title} {\bibinfo {title} {Testing quantum
  nonlocality in phase space},\ }\href
  {https://doi.org/10.1103/PhysRevLett.82.2009} {\bibfield  {journal} {\bibinfo
   {journal} {Phys. Rev. Lett.}\ }\textbf {\bibinfo {volume} {82}},\ \bibinfo
  {pages} {2009} (\bibinfo {year} {1999})}\BibitemShut {NoStop}%
\bibitem [{\citenamefont {Wenger}\ \emph {et~al.}(2003)\citenamefont {Wenger},
  \citenamefont {Hafezi}, \citenamefont {Grosshans}, \citenamefont
  {Tualle-Brouri},\ and\ \citenamefont {Grangier}}]{wenger03}%
  \BibitemOpen
  \bibfield  {author} {\bibinfo {author} {\bibfnamefont {J.}~\bibnamefont
  {Wenger}}, \bibinfo {author} {\bibfnamefont {M.}~\bibnamefont {Hafezi}},
  \bibinfo {author} {\bibfnamefont {F.}~\bibnamefont {Grosshans}}, \bibinfo
  {author} {\bibfnamefont {R.}~\bibnamefont {Tualle-Brouri}},\ and\ \bibinfo
  {author} {\bibfnamefont {P.}~\bibnamefont {Grangier}},\ }\bibfield  {title}
  {\bibinfo {title} {Maximal violation of bell inequalities using
  continuous-variable measurements},\ }\href
  {https://doi.org/10.1103/PhysRevA.67.012105} {\bibfield  {journal} {\bibinfo
  {journal} {Phys. Rev. A}\ }\textbf {\bibinfo {volume} {67}},\ \bibinfo
  {pages} {012105} (\bibinfo {year} {2003})}\BibitemShut {NoStop}%
\bibitem [{\citenamefont {Cavalcanti}\ \emph {et~al.}(2011)\citenamefont
  {Cavalcanti}, \citenamefont {Brunner}, \citenamefont {Skrzypczyk},
  \citenamefont {Salles},\ and\ \citenamefont {Scarani}}]{cavalcanti11}%
  \BibitemOpen
  \bibfield  {author} {\bibinfo {author} {\bibfnamefont {D.}~\bibnamefont
  {Cavalcanti}}, \bibinfo {author} {\bibfnamefont {N.}~\bibnamefont {Brunner}},
  \bibinfo {author} {\bibfnamefont {P.}~\bibnamefont {Skrzypczyk}}, \bibinfo
  {author} {\bibfnamefont {A.}~\bibnamefont {Salles}},\ and\ \bibinfo {author}
  {\bibfnamefont {V.}~\bibnamefont {Scarani}},\ }\bibfield  {title} {\bibinfo
  {title} {Large violation of bell inequalities using both particle andwave
  measurements},\ }\href {https://doi.org/10.1103/PhysRevA.84.022105}
  {\bibfield  {journal} {\bibinfo  {journal} {Phys. Rev. A}\ }\textbf {\bibinfo
  {volume} {84}},\ \bibinfo {pages} {022105} (\bibinfo {year}
  {2011})}\BibitemShut {NoStop}%
\bibitem [{\citenamefont {Ketterer}\ \emph {et~al.}(2015)\citenamefont
  {Ketterer}, \citenamefont {Keller}, \citenamefont {Coudreau},\ and\
  \citenamefont {Milman}}]{ketterer15}%
  \BibitemOpen
  \bibfield  {author} {\bibinfo {author} {\bibfnamefont {A.}~\bibnamefont
  {Ketterer}}, \bibinfo {author} {\bibfnamefont {A.}~\bibnamefont {Keller}},
  \bibinfo {author} {\bibfnamefont {T.}~\bibnamefont {Coudreau}},\ and\
  \bibinfo {author} {\bibfnamefont {P.}~\bibnamefont {Milman}},\ }\bibfield
  {title} {\bibinfo {title} {Testing the clauser-horne-shimony-holt inequality
  using observables with arbitrary spectrum},\ }\href
  {https://doi.org/10.1103/PhysRevA.91.012106} {\bibfield  {journal} {\bibinfo
  {journal} {Phys. Rev. A}\ }\textbf {\bibinfo {volume} {91}},\ \bibinfo
  {pages} {012106} (\bibinfo {year} {2015})}\BibitemShut {NoStop}%
\bibitem [{\citenamefont {Massar}\ and\ \citenamefont
  {Pironio}(2001)}]{massar01}%
  \BibitemOpen
  \bibfield  {author} {\bibinfo {author} {\bibfnamefont {S.}~\bibnamefont
  {Massar}}\ and\ \bibinfo {author} {\bibfnamefont {S.}~\bibnamefont
  {Pironio}},\ }\bibfield  {title} {\bibinfo {title}
  {Greenberger-horne-zeilinger paradox for continuous variables},\ }\href
  {https://doi.org/10.1103/PhysRevA.64.062108} {\bibfield  {journal} {\bibinfo
  {journal} {Phys. Rev. A}\ }\textbf {\bibinfo {volume} {64}},\ \bibinfo
  {pages} {062108} (\bibinfo {year} {2001})}\BibitemShut {NoStop}%
\bibitem [{\citenamefont {Plastino}\ and\ \citenamefont
  {Cabello}(2010)}]{plastino10}%
  \BibitemOpen
  \bibfield  {author} {\bibinfo {author} {\bibfnamefont {A.~R.}\ \bibnamefont
  {Plastino}}\ and\ \bibinfo {author} {\bibfnamefont {A.}~\bibnamefont
  {Cabello}},\ }\bibfield  {title} {\bibinfo {title} {State-independent quantum
  contextuality for continuous variables},\ }\href@noop {} {\bibfield
  {journal} {\bibinfo  {journal} {Phys. Rev. A}\ }\textbf {\bibinfo {volume}
  {82}},\ \bibinfo {pages} {022114} (\bibinfo {year} {2010})}\BibitemShut
  {NoStop}%
\bibitem [{\citenamefont {Asadian}\ \emph {et~al.}(2015)\citenamefont
  {Asadian}, \citenamefont {Budroni}, \citenamefont {Steinhoff}, \citenamefont
  {Rabl},\ and\ \citenamefont {G\"uhne}}]{asadian15}%
  \BibitemOpen
  \bibfield  {author} {\bibinfo {author} {\bibfnamefont {A.}~\bibnamefont
  {Asadian}}, \bibinfo {author} {\bibfnamefont {C.}~\bibnamefont {Budroni}},
  \bibinfo {author} {\bibfnamefont {F.~E.~S.}\ \bibnamefont {Steinhoff}},
  \bibinfo {author} {\bibfnamefont {P.}~\bibnamefont {Rabl}},\ and\ \bibinfo
  {author} {\bibfnamefont {O.}~\bibnamefont {G\"uhne}},\ }\bibfield  {title}
  {\bibinfo {title} {Contextuality in phase space},\ }\href
  {https://doi.org/10.1103/PhysRevLett.114.250403} {\bibfield  {journal}
  {\bibinfo  {journal} {Phys. Rev. Lett.}\ }\textbf {\bibinfo {volume} {114}},\
  \bibinfo {pages} {250403} (\bibinfo {year} {2015})}\BibitemShut {NoStop}%
\bibitem [{\citenamefont {Laversanne-Finot}\ \emph {et~al.}(2017)\citenamefont
  {Laversanne-Finot}, \citenamefont {Ketterer}, \citenamefont {Barros},
  \citenamefont {Walborn}, \citenamefont {Coudreau}, \citenamefont {Keller},\
  and\ \citenamefont {Milman}}]{finot17}%
  \BibitemOpen
  \bibfield  {author} {\bibinfo {author} {\bibfnamefont {A.}~\bibnamefont
  {Laversanne-Finot}}, \bibinfo {author} {\bibfnamefont {A.}~\bibnamefont
  {Ketterer}}, \bibinfo {author} {\bibfnamefont {M.~R.}\ \bibnamefont
  {Barros}}, \bibinfo {author} {\bibfnamefont {S.~P.}\ \bibnamefont {Walborn}},
  \bibinfo {author} {\bibfnamefont {T.}~\bibnamefont {Coudreau}}, \bibinfo
  {author} {\bibfnamefont {A.}~\bibnamefont {Keller}},\ and\ \bibinfo {author}
  {\bibfnamefont {P.}~\bibnamefont {Milman}},\ }\bibfield  {title} {\bibinfo
  {title} {General conditions for maximal violation of non-contextuality in
  discrete and continuous variables},\ }\href@noop {} {\bibfield  {journal}
  {\bibinfo  {journal} {Journal of Physics A: Mathematical and Theoretical}\
  }\textbf {\bibinfo {volume} {50}},\ \bibinfo {pages} {155304} (\bibinfo
  {year} {2017})}\BibitemShut {NoStop}%
\bibitem [{\citenamefont {Ketterer}\ \emph {et~al.}(2014)\citenamefont
  {Ketterer}, \citenamefont {Douce}, \citenamefont {Keller}, \citenamefont
  {Coudreau},\ and\ \citenamefont {Milman}}]{ketterer14}%
  \BibitemOpen
  \bibfield  {author} {\bibinfo {author} {\bibfnamefont {A.}~\bibnamefont
  {Ketterer}}, \bibinfo {author} {\bibfnamefont {T.}~\bibnamefont {Douce}},
  \bibinfo {author} {\bibfnamefont {A.}~\bibnamefont {Keller}}, \bibinfo
  {author} {\bibfnamefont {T.}~\bibnamefont {Coudreau}},\ and\ \bibinfo
  {author} {\bibfnamefont {P.}~\bibnamefont {Milman}},\ }\href@noop {}
  {\bibinfo {title} {Quantum search with modular variables}} (\bibinfo {year}
  {2014}),\ \Eprint {https://arxiv.org/abs/1407.1298} {arXiv:1407.1298
  [quant-ph]} \BibitemShut {NoStop}%
\bibitem [{\citenamefont {Gneiting}\ and\ \citenamefont
  {Hornberger}(2011)}]{Gneiting11}%
  \BibitemOpen
  \bibfield  {author} {\bibinfo {author} {\bibfnamefont {C.}~\bibnamefont
  {Gneiting}}\ and\ \bibinfo {author} {\bibfnamefont {K.}~\bibnamefont
  {Hornberger}},\ }\bibfield  {title} {\bibinfo {title} {Detecting entanglement
  in spatial interference},\ }\href@noop {} {\bibfield  {journal} {\bibinfo
  {journal} {Phys. Rev. Lett.}\ }\textbf {\bibinfo {volume} {106}},\ \bibinfo
  {pages} {210501} (\bibinfo {year} {2011})}\BibitemShut {NoStop}%
\bibitem [{\citenamefont {Carvalho}\ \emph {et~al.}(2012)\citenamefont
  {Carvalho}, \citenamefont {Ferraz}, \citenamefont {Borges}, \citenamefont
  {de~Assis}, \citenamefont {P{\'a}dua},\ and\ \citenamefont
  {Walborn}}]{Carvalho12}%
  \BibitemOpen
  \bibfield  {author} {\bibinfo {author} {\bibfnamefont {M.~A.~D.}\
  \bibnamefont {Carvalho}}, \bibinfo {author} {\bibfnamefont {J.}~\bibnamefont
  {Ferraz}}, \bibinfo {author} {\bibfnamefont {G.~F.}\ \bibnamefont {Borges}},
  \bibinfo {author} {\bibfnamefont {P.-L.}\ \bibnamefont {de~Assis}}, \bibinfo
  {author} {\bibfnamefont {S.}~\bibnamefont {P{\'a}dua}},\ and\ \bibinfo
  {author} {\bibfnamefont {S.~P.}\ \bibnamefont {Walborn}},\ }\bibfield
  {title} {\bibinfo {title} {Experimental observation of quantum correlations
  in modular variables},\ }\href@noop {} {\bibfield  {journal} {\bibinfo
  {journal} {Phys. Rev. A}\ }\textbf {\bibinfo {volume} {86}},\ \bibinfo
  {pages} {032332} (\bibinfo {year} {2012})}\BibitemShut {NoStop}%
\bibitem [{\citenamefont {Tasca}\ \emph
  {et~al.}(2018{\natexlab{a}})\citenamefont {Tasca}, \citenamefont {Rudnicki},
  \citenamefont {Aspden}, \citenamefont {Padgett}, \citenamefont
  {Souto~Ribeiro},\ and\ \citenamefont {Walborn}}]{Tasca18b}%
  \BibitemOpen
  \bibfield  {author} {\bibinfo {author} {\bibfnamefont {D.~S.}\ \bibnamefont
  {Tasca}}, \bibinfo {author} {\bibfnamefont {{\L}.}~\bibnamefont {Rudnicki}},
  \bibinfo {author} {\bibfnamefont {R.~S.}\ \bibnamefont {Aspden}}, \bibinfo
  {author} {\bibfnamefont {M.~J.}\ \bibnamefont {Padgett}}, \bibinfo {author}
  {\bibfnamefont {P.~H.}\ \bibnamefont {Souto~Ribeiro}},\ and\ \bibinfo
  {author} {\bibfnamefont {S.~P.}\ \bibnamefont {Walborn}},\ }\bibfield
  {title} {\bibinfo {title} {Testing for entanglement with periodic coarse
  graining},\ }\href {https://doi.org/10.1103/PhysRevA.97.042312} {\bibfield
  {journal} {\bibinfo  {journal} {Phys. Rev. A}\ }\textbf {\bibinfo {volume}
  {97}},\ \bibinfo {pages} {042312} (\bibinfo {year}
  {2018}{\natexlab{a}})}\BibitemShut {NoStop}%
\bibitem [{\citenamefont {Vernaz-Gris}\ \emph {et~al.}(2014)\citenamefont
  {Vernaz-Gris}, \citenamefont {Ketterer}, \citenamefont {Keller},
  \citenamefont {Walborn}, \citenamefont {Coudreau},\ and\ \citenamefont
  {Milman}}]{Vernaz-Gris14}%
  \BibitemOpen
  \bibfield  {author} {\bibinfo {author} {\bibfnamefont {P.}~\bibnamefont
  {Vernaz-Gris}}, \bibinfo {author} {\bibfnamefont {A.}~\bibnamefont
  {Ketterer}}, \bibinfo {author} {\bibfnamefont {A.}~\bibnamefont {Keller}},
  \bibinfo {author} {\bibfnamefont {S.~P.}\ \bibnamefont {Walborn}}, \bibinfo
  {author} {\bibfnamefont {T.}~\bibnamefont {Coudreau}},\ and\ \bibinfo
  {author} {\bibfnamefont {P.}~\bibnamefont {Milman}},\ }\bibfield  {title}
  {\bibinfo {title} {Continuous discretization of infinite-dimensional hilbert
  spaces},\ }\href@noop {} {\bibfield  {journal} {\bibinfo  {journal} {Phys.
  Rev. A}\ }\textbf {\bibinfo {volume} {89}},\ \bibinfo {pages} {052311}
  (\bibinfo {year} {2014})}\BibitemShut {NoStop}%
\bibitem [{\citenamefont {Ketterer}\ \emph {et~al.}(2016)\citenamefont
  {Ketterer}, \citenamefont {Keller}, \citenamefont {Walborn}, \citenamefont
  {Coudreau},\ and\ \citenamefont {Milman}}]{Ketterer16}%
  \BibitemOpen
  \bibfield  {author} {\bibinfo {author} {\bibfnamefont {A.}~\bibnamefont
  {Ketterer}}, \bibinfo {author} {\bibfnamefont {A.}~\bibnamefont {Keller}},
  \bibinfo {author} {\bibfnamefont {S.~P.}\ \bibnamefont {Walborn}}, \bibinfo
  {author} {\bibfnamefont {T.}~\bibnamefont {Coudreau}},\ and\ \bibinfo
  {author} {\bibfnamefont {P.}~\bibnamefont {Milman}},\ }\bibfield  {title}
  {\bibinfo {title} {Quantum information processing in phase space: A modular
  variables approach},\ }\href {https://doi.org/10.1103/PhysRevA.94.022325}
  {\bibfield  {journal} {\bibinfo  {journal} {Phys. Rev. A}\ }\textbf {\bibinfo
  {volume} {94}},\ \bibinfo {pages} {022325} (\bibinfo {year}
  {2016})}\BibitemShut {NoStop}%
\bibitem [{\citenamefont {Weigert}\ and\ \citenamefont
  {Wilkinson}(2008)}]{Weigert08}%
  \BibitemOpen
  \bibfield  {author} {\bibinfo {author} {\bibfnamefont {S.}~\bibnamefont
  {Weigert}}\ and\ \bibinfo {author} {\bibfnamefont {M.}~\bibnamefont
  {Wilkinson}},\ }\bibfield  {title} {\bibinfo {title} {Mutually unbiased bases
  for continuous variables},\ }\href@noop {} {\bibfield  {journal} {\bibinfo
  {journal} {Phys. Rev. A}\ }\textbf {\bibinfo {volume} {78}},\ \bibinfo
  {pages} {020303(R)} (\bibinfo {year} {2008})}\BibitemShut {NoStop}%
\bibitem [{\citenamefont {Ivonovic}(1981)}]{Ivonovic1981}%
  \BibitemOpen
  \bibfield  {author} {\bibinfo {author} {\bibfnamefont {I.~D.}\ \bibnamefont
  {Ivonovic}},\ }\bibfield  {title} {\bibinfo {title} {Geometrical description
  of quantal state determination},\ }\href
  {https://doi.org/10.1088/0305-4470/14/12/019} {\bibfield  {journal} {\bibinfo
   {journal} {Journal of Physics A: Mathematical and General}\ }\textbf
  {\bibinfo {volume} {14}},\ \bibinfo {pages} {3241} (\bibinfo {year}
  {1981})}\BibitemShut {NoStop}%
\bibitem [{\citenamefont {Wootters}\ and\ \citenamefont
  {Fields}(1989)}]{wootters1989}%
  \BibitemOpen
  \bibfield  {author} {\bibinfo {author} {\bibfnamefont {W.~K.}\ \bibnamefont
  {Wootters}}\ and\ \bibinfo {author} {\bibfnamefont {B.~D.}\ \bibnamefont
  {Fields}},\ }\bibfield  {title} {\bibinfo {title} {Optimal
  state-determination by mutually unbiased measurements},\ }\href@noop {}
  {\bibfield  {journal} {\bibinfo  {journal} {Annals of Physics}\ }\textbf
  {\bibinfo {volume} {191}},\ \bibinfo {pages} {363} (\bibinfo {year}
  {1989})}\BibitemShut {NoStop}%
\bibitem [{\citenamefont {Bandyopadhyay}\ \emph {et~al.}(2002)\citenamefont
  {Bandyopadhyay}, \citenamefont {Boykin}, \citenamefont {Roychowdhury},\ and\
  \citenamefont {Vatan}}]{Bandyopadhyay2002}%
  \BibitemOpen
  \bibfield  {author} {\bibinfo {author} {\bibfnamefont {S.}~\bibnamefont
  {Bandyopadhyay}}, \bibinfo {author} {\bibfnamefont {P.}~\bibnamefont
  {Boykin}}, \bibinfo {author} {\bibfnamefont {V.}~\bibnamefont
  {Roychowdhury}},\ and\ \bibinfo {author} {\bibfnamefont {F.}~\bibnamefont
  {Vatan}},\ }\bibfield  {title} {\bibinfo {title} {A new proof ofthe existence
  of mutually unbiased bases},\ }\href@noop {} {\bibfield  {journal} {\bibinfo
  {journal} {Algorithmica}\ }\textbf {\bibinfo {volume} {34}},\ \bibinfo
  {pages} {512} (\bibinfo {year} {2002})}\BibitemShut {NoStop}%
\bibitem [{\citenamefont {Klappenecker}\ and\ \citenamefont
  {R{\"o}tteler}(2003)}]{klappenecker2003}%
  \BibitemOpen
  \bibfield  {author} {\bibinfo {author} {\bibfnamefont {A.}~\bibnamefont
  {Klappenecker}}\ and\ \bibinfo {author} {\bibfnamefont {M.}~\bibnamefont
  {R{\"o}tteler}},\ }\bibfield  {title} {\bibinfo {title} {Constructions of
  mutually unbiased bases},\ }in\ \href@noop {} {\emph {\bibinfo {booktitle}
  {Finite Fields and Applications}}},\ \bibinfo {editor} {edited by\ \bibinfo
  {editor} {\bibfnamefont {G.}~\bibnamefont {Mullen}}, \bibinfo {editor}
  {\bibfnamefont {A.}~\bibnamefont {Poli}},\ and\ \bibinfo {editor}
  {\bibfnamefont {H.}~\bibnamefont {Stichtenoth}}}\ (\bibinfo  {publisher}
  {Springer},\ \bibinfo {year} {2003})\ p.\ \bibinfo {pages} {137}\BibitemShut
  {NoStop}%
\bibitem [{\citenamefont {Schwinger}(1960)}]{Schwinger60}%
  \BibitemOpen
  \bibfield  {author} {\bibinfo {author} {\bibfnamefont {J.}~\bibnamefont
  {Schwinger}},\ }\bibfield  {title} {\bibinfo {title} {Unitary operator
  bases},\ }\href@noop {} {\bibfield  {journal} {\bibinfo  {journal} {Proc.
  Natl. Acad. Sci.}\ }\textbf {\bibinfo {volume} {46}},\ \bibinfo {pages} {570}
  (\bibinfo {year} {1960})}\BibitemShut {NoStop}%
\bibitem [{\citenamefont {Kraus}(1987)}]{Kraus87}%
  \BibitemOpen
  \bibfield  {author} {\bibinfo {author} {\bibfnamefont {K.}~\bibnamefont
  {Kraus}},\ }\bibfield  {title} {\bibinfo {title} {Complementary observables
  and uncertainty relations},\ }\href@noop {} {\bibfield  {journal} {\bibinfo
  {journal} {Phys. Rev. D}\ }\textbf {\bibinfo {volume} {35}},\ \bibinfo
  {pages} {3070} (\bibinfo {year} {1987})}\BibitemShut {NoStop}%
\bibitem [{\citenamefont {Durt}\ \emph {et~al.}(2010)\citenamefont {Durt},
  \citenamefont {Englert}, \citenamefont {Bengtsson},\ and\ \citenamefont
  {{\.Z}yczkowski}}]{Durt10}%
  \BibitemOpen
  \bibfield  {author} {\bibinfo {author} {\bibfnamefont {T.}~\bibnamefont
  {Durt}}, \bibinfo {author} {\bibfnamefont {B.-G.}\ \bibnamefont {Englert}},
  \bibinfo {author} {\bibfnamefont {I.}~\bibnamefont {Bengtsson}},\ and\
  \bibinfo {author} {\bibfnamefont {K.}~\bibnamefont {{\.Z}yczkowski}},\
  }\bibfield  {title} {\bibinfo {title} {On mutually unbiased bases},\
  }\href@noop {} {\bibfield  {journal} {\bibinfo  {journal} {Int. J. Quant.
  Inf.}\ }\textbf {\bibinfo {volume} {08}},\ \bibinfo {pages} {535} (\bibinfo
  {year} {2010})}\BibitemShut {NoStop}%
\bibitem [{Note1()}]{Note1}%
  \BibitemOpen
  \bibinfo {note} {We note that in the limit $\theta _{jk} \rightarrow 0$, the
  limit must be taken before the absolute value to recover the normalization to
  the usual Dirac delta function.}\BibitemShut {Stop}%
\bibitem [{\citenamefont {Tasca}\ \emph {et~al.}(2013)\citenamefont {Tasca},
  \citenamefont {Rudnicki}, \citenamefont {Gomes}, \citenamefont {Toscano},\
  and\ \citenamefont {Walborn}}]{tasca13}%
  \BibitemOpen
  \bibfield  {author} {\bibinfo {author} {\bibfnamefont {D.~S.}\ \bibnamefont
  {Tasca}}, \bibinfo {author} {\bibfnamefont {{\L}.}~\bibnamefont {Rudnicki}},
  \bibinfo {author} {\bibfnamefont {R.~M.}\ \bibnamefont {Gomes}}, \bibinfo
  {author} {\bibfnamefont {F.}~\bibnamefont {Toscano}},\ and\ \bibinfo {author}
  {\bibfnamefont {S.~P.}\ \bibnamefont {Walborn}},\ }\bibfield  {title}
  {\bibinfo {title} {Reliable entanglement detection under coarse-grained
  measurements},\ }\href@noop {} {\bibfield  {journal} {\bibinfo  {journal}
  {Phys. Rev. Lett.}\ }\textbf {\bibinfo {volume} {110}},\ \bibinfo {pages}
  {210502} (\bibinfo {year} {2013})}\BibitemShut {NoStop}%
\bibitem [{\citenamefont {Ray}\ and\ \citenamefont {van
  Enk}(2013{\natexlab{a}})}]{Ray13a}%
  \BibitemOpen
  \bibfield  {author} {\bibinfo {author} {\bibfnamefont {M.~R.}\ \bibnamefont
  {Ray}}\ and\ \bibinfo {author} {\bibfnamefont {S.~J.}\ \bibnamefont {van
  Enk}},\ }\bibfield  {title} {\bibinfo {title} {Missing data outside the
  detector range: Continuous-variable entanglement verification and quantum
  cryptography},\ }\href@noop {} {\bibfield  {journal} {\bibinfo  {journal}
  {Phys. Rev. A}\ }\textbf {\bibinfo {volume} {88}},\ \bibinfo {pages} {042326}
  (\bibinfo {year} {2013}{\natexlab{a}})}\BibitemShut {NoStop}%
\bibitem [{\citenamefont {Ray}\ and\ \citenamefont {van
  Enk}(2013{\natexlab{b}})}]{Ray13b}%
  \BibitemOpen
  \bibfield  {author} {\bibinfo {author} {\bibfnamefont {M.~R.}\ \bibnamefont
  {Ray}}\ and\ \bibinfo {author} {\bibfnamefont {S.~J.}\ \bibnamefont {van
  Enk}},\ }\bibfield  {title} {\bibinfo {title} {Missing data outside the
  detector range. ii. application to time-frequency entanglement},\ }\href@noop
  {} {\bibfield  {journal} {\bibinfo  {journal} {Phys. Rev. A}\ }\textbf
  {\bibinfo {volume} {88}},\ \bibinfo {pages} {062327} (\bibinfo {year}
  {2013}{\natexlab{b}})}\BibitemShut {NoStop}%
\bibitem [{\citenamefont {Kalev}\ and\ \citenamefont {Gour}(2014)}]{Kalev14}%
  \BibitemOpen
  \bibfield  {author} {\bibinfo {author} {\bibfnamefont {A.}~\bibnamefont
  {Kalev}}\ and\ \bibinfo {author} {\bibfnamefont {G.}~\bibnamefont {Gour}},\
  }\bibfield  {title} {\bibinfo {title} {Mutually unbiased measurements in
  finite dimensions},\ }\href@noop {} {\bibfield  {journal} {\bibinfo
  {journal} {New J. Phys.}\ }\textbf {\bibinfo {volume} {16}},\ \bibinfo
  {pages} {053038} (\bibinfo {year} {2014})}\BibitemShut {NoStop}%
\bibitem [{\citenamefont {Tasca}\ \emph
  {et~al.}(2018{\natexlab{b}})\citenamefont {Tasca}, \citenamefont
  {S{\'a}nchez}, \citenamefont {Walborn},\ and\ \citenamefont
  {Rudnicki}}]{Tasca18a}%
  \BibitemOpen
  \bibfield  {author} {\bibinfo {author} {\bibfnamefont {D.~S.}\ \bibnamefont
  {Tasca}}, \bibinfo {author} {\bibfnamefont {P.}~\bibnamefont {S{\'a}nchez}},
  \bibinfo {author} {\bibfnamefont {S.~P.}\ \bibnamefont {Walborn}},\ and\
  \bibinfo {author} {\bibfnamefont {{\L}.}~\bibnamefont {Rudnicki}},\
  }\bibfield  {title} {\bibinfo {title} {Mutual unbiasedness in coarse-grained
  continuous variables},\ }\href@noop {} {\bibfield  {journal} {\bibinfo
  {journal} {Phys. Rev. Lett.}\ }\textbf {\bibinfo {volume} {120}},\ \bibinfo
  {pages} {040403} (\bibinfo {year} {2018}{\natexlab{b}})}\BibitemShut
  {NoStop}%
\bibitem [{\citenamefont {Paul}\ \emph {et~al.}(2018)\citenamefont {Paul},
  \citenamefont {Walborn}, \citenamefont {Tasca},\ and\ \citenamefont
  {Rudnicki}}]{paul18}%
  \BibitemOpen
  \bibfield  {author} {\bibinfo {author} {\bibfnamefont {E.~C.}\ \bibnamefont
  {Paul}}, \bibinfo {author} {\bibfnamefont {S.~P.}\ \bibnamefont {Walborn}},
  \bibinfo {author} {\bibfnamefont {D.~S.}\ \bibnamefont {Tasca}},\ and\
  \bibinfo {author} {\bibfnamefont {{\L}.}~\bibnamefont {Rudnicki}},\
  }\bibfield  {title} {\bibinfo {title} {Mutually unbiased coarse-grained
  measurements of two or more phase-space variables},\ }\href@noop {}
  {\bibfield  {journal} {\bibinfo  {journal} {Phys. Rev. A}\ }\textbf {\bibinfo
  {volume} {97}},\ \bibinfo {pages} {052103} (\bibinfo {year}
  {2018})}\BibitemShut {NoStop}%
\bibitem [{\citenamefont {Tavakoli}\ \emph {et~al.}(2019)\citenamefont
  {Tavakoli}, \citenamefont {Farkas}, \citenamefont {Rosset}, \citenamefont
  {Bancal},\ and\ \citenamefont {Kaniewski}}]{kaniewski19}%
  \BibitemOpen
  \bibfield  {author} {\bibinfo {author} {\bibfnamefont {A.}~\bibnamefont
  {Tavakoli}}, \bibinfo {author} {\bibfnamefont {M.}~\bibnamefont {Farkas}},
  \bibinfo {author} {\bibfnamefont {D.}~\bibnamefont {Rosset}}, \bibinfo
  {author} {\bibfnamefont {J.-D.}\ \bibnamefont {Bancal}},\ and\ \bibinfo
  {author} {\bibfnamefont {J.}~\bibnamefont {Kaniewski}},\ }\bibfield  {title}
  {\bibinfo {title} {Mutually unbiased bases and symmetric informationally
  complete measurements in bell experiments: Bell inequalities,
  device-independent certification and applications},\ }\href@noop {}
  {\bibfield  {journal} {\bibinfo  {journal} {arXiv:1912.03225}\ } (\bibinfo
  {year} {2019})}\BibitemShut {NoStop}%
\bibitem [{\citenamefont {Marcuse}(1982)}]{marcuse82}%
  \BibitemOpen
  \bibinfo {editor} {\bibfnamefont {D.}~\bibnamefont {Marcuse}},\ ed.,\
  \href@noop {} {\emph {\bibinfo {title} {Light Transmission Optics}}}\
  (\bibinfo  {publisher} {Van Nostrand Reinhold Publishing},\ \bibinfo
  {address} {New York},\ \bibinfo {year} {1982})\BibitemShut {NoStop}%
\bibitem [{\citenamefont {Tasca}\ \emph {et~al.}(2011)\citenamefont {Tasca},
  \citenamefont {Gomes}, \citenamefont {Toscano}, \citenamefont
  {Souto~Ribeiro},\ and\ \citenamefont {Walborn}}]{tasca11}%
  \BibitemOpen
  \bibfield  {author} {\bibinfo {author} {\bibfnamefont {D.~S.}\ \bibnamefont
  {Tasca}}, \bibinfo {author} {\bibfnamefont {R.~M.}\ \bibnamefont {Gomes}},
  \bibinfo {author} {\bibfnamefont {F.}~\bibnamefont {Toscano}}, \bibinfo
  {author} {\bibfnamefont {P.~H.}\ \bibnamefont {Souto~Ribeiro}},\ and\
  \bibinfo {author} {\bibfnamefont {S.~P.}\ \bibnamefont {Walborn}},\
  }\bibfield  {title} {\bibinfo {title} {Continuous-variable quantum
  computation with spatial degrees of freedom of photons},\ }\href
  {https://doi.org/10.1103/PhysRevA.83.052325} {\bibfield  {journal} {\bibinfo
  {journal} {Phys. Rev. A}\ }\textbf {\bibinfo {volume} {83}},\ \bibinfo
  {pages} {052325} (\bibinfo {year} {2011})}\BibitemShut {NoStop}%
\bibitem [{\citenamefont {Ozaktas}\ \emph {et~al.}(2001)\citenamefont
  {Ozaktas}, \citenamefont {Zalevsky},\ and\ \citenamefont
  {Kutay}}]{ozaktas01}%
  \BibitemOpen
  \bibfield  {author} {\bibinfo {author} {\bibfnamefont {H.~M.}\ \bibnamefont
  {Ozaktas}}, \bibinfo {author} {\bibfnamefont {Z.}~\bibnamefont {Zalevsky}},\
  and\ \bibinfo {author} {\bibfnamefont {M.~A.}\ \bibnamefont {Kutay}},\
  }\href@noop {} {\emph {\bibinfo {title} {The Fractional Fourier Transform:
  with Applications in Optics and Signal Processing}}}\ (\bibinfo  {publisher}
  {John Wiley and Sons Ltd},\ \bibinfo {address} {New York},\ \bibinfo {year}
  {2001})\BibitemShut {NoStop}%
\bibitem [{\citenamefont {Tasca}\ \emph {et~al.}(2008)\citenamefont {Tasca},
  \citenamefont {Walborn}, \citenamefont {Souto~Ribeiro},\ and\ \citenamefont
  {Toscano}}]{tasca08}%
  \BibitemOpen
  \bibfield  {author} {\bibinfo {author} {\bibfnamefont {D.~S.}\ \bibnamefont
  {Tasca}}, \bibinfo {author} {\bibfnamefont {S.~P.}\ \bibnamefont {Walborn}},
  \bibinfo {author} {\bibfnamefont {P.~H.}\ \bibnamefont {Souto~Ribeiro}},\
  and\ \bibinfo {author} {\bibfnamefont {F.}~\bibnamefont {Toscano}},\
  }\bibfield  {title} {\bibinfo {title} {Detection of transverse entanglement
  in phase space},\ }\href {https://doi.org/10.1103/PhysRevA.78.010304}
  {\bibfield  {journal} {\bibinfo  {journal} {Phys. Rev. A}\ }\textbf {\bibinfo
  {volume} {78}},\ \bibinfo {pages} {010304(R)} (\bibinfo {year}
  {2008})}\BibitemShut {NoStop}%
\bibitem [{\citenamefont {Tasca}\ \emph {et~al.}(2009)\citenamefont {Tasca},
  \citenamefont {Walborn}, \citenamefont {Souto~Ribeiro}, \citenamefont
  {Toscano},\ and\ \citenamefont {Pellat-Finet}}]{tasca09a}%
  \BibitemOpen
  \bibfield  {author} {\bibinfo {author} {\bibfnamefont {D.~S.}\ \bibnamefont
  {Tasca}}, \bibinfo {author} {\bibfnamefont {S.~P.}\ \bibnamefont {Walborn}},
  \bibinfo {author} {\bibfnamefont {P.~H.}\ \bibnamefont {Souto~Ribeiro}},
  \bibinfo {author} {\bibfnamefont {F.}~\bibnamefont {Toscano}},\ and\ \bibinfo
  {author} {\bibfnamefont {P.}~\bibnamefont {Pellat-Finet}},\ }\bibfield
  {title} {\bibinfo {title} {Propagation of transverse intensity correlations
  of a two-photon state},\ }\href {https://doi.org/10.1103/PhysRevA.79.033801}
  {\bibfield  {journal} {\bibinfo  {journal} {Phys. Rev. A}\ }\textbf {\bibinfo
  {volume} {79}},\ \bibinfo {pages} {033801} (\bibinfo {year}
  {2009})}\BibitemShut {NoStop}%
\bibitem [{\citenamefont {Paul}\ \emph {et~al.}(2016)\citenamefont {Paul},
  \citenamefont {Tasca}, \citenamefont {Rudnicki},\ and\ \citenamefont
  {Walborn}}]{Paul16}%
  \BibitemOpen
  \bibfield  {author} {\bibinfo {author} {\bibfnamefont {E.~C.}\ \bibnamefont
  {Paul}}, \bibinfo {author} {\bibfnamefont {D.~S.}\ \bibnamefont {Tasca}},
  \bibinfo {author} {\bibfnamefont {{\L}.}~\bibnamefont {Rudnicki}},\ and\
  \bibinfo {author} {\bibfnamefont {S.~P.}\ \bibnamefont {Walborn}},\
  }\bibfield  {title} {\bibinfo {title} {Detecting entanglement of continuous
  variables with three mutually unbiased bases},\ }\href@noop {} {\bibfield
  {journal} {\bibinfo  {journal} {Phys. Rev. A}\ }\textbf {\bibinfo {volume}
  {94}},\ \bibinfo {pages} {012303} (\bibinfo {year} {2016})}\BibitemShut
  {NoStop}%
\bibitem [{\citenamefont {Rodrigo}\ \emph {et~al.}(2009)\citenamefont
  {Rodrigo}, \citenamefont {Alieva},\ and\ \citenamefont {Calvo}}]{Rodrigo09}%
  \BibitemOpen
  \bibfield  {author} {\bibinfo {author} {\bibfnamefont {J.~A.}\ \bibnamefont
  {Rodrigo}}, \bibinfo {author} {\bibfnamefont {T.}~\bibnamefont {Alieva}},\
  and\ \bibinfo {author} {\bibfnamefont {M.~L.}\ \bibnamefont {Calvo}},\
  }\bibfield  {title} {\bibinfo {title} {Programmable two-dimensional optical
  fractional fourier processor},\ }\href {https://doi.org/10.1364/OE.17.004976}
  {\bibfield  {journal} {\bibinfo  {journal} {Opt. Express}\ }\textbf {\bibinfo
  {volume} {17}},\ \bibinfo {pages} {4976} (\bibinfo {year}
  {2009})}\BibitemShut {NoStop}%
\end{thebibliography}%

\end{document}